\documentclass[a4paper,10pt]{article}
\usepackage{newlfont}
\usepackage[T1]{fontenc}
\usepackage[utf8]{inputenc}
\usepackage{indentfirst}
\usepackage{fancyhdr}
\usepackage{graphicx} 
\usepackage{amssymb}
\usepackage{amsmath} 
\usepackage{latexsym}
\usepackage{amsthm} 
\usepackage{float}

\usepackage[english]{babel}
\usepackage{amsthm}
\usepackage{wrapfig}
\usepackage{subfigure}

\theoremstyle{plain} 
\newtheorem{thm}{Theorem}[section] 

\newtheorem{lem}[thm]{Lemma}

\theoremstyle{definition} 
\newtheorem{defn}{Definition}[section]

\theoremstyle{remark} 
\newtheorem{rem}{Remark}[section]
\newtheorem{ex}{Example}[section]

\title{Inverse problem for the mean-field monomer-dimer model with attractive interaction}
\author{Pierluigi Contucci, Rachele Luzi, Cecilia Vernia}
\date{ }
\begin{document}
\maketitle
\begin{abstract}
The inverse problem method is tested for a class of monomer-dimer statistical mechanics models 
that contain also an attractive potential and display a mean-field critical point at a boundary of a coexistence line. 
The inversion is obtained by analytically identifying the parameters in terms
of the correlation functions and via the maximum-likelihood method. The precision is tested in the whole phase space
and, when close to the coexistence line, the algorithm is used together with a clustering method to take care of the underlying 
possible ambiguity of the inversion.
\end{abstract}

\section{Introduction}

In the last decade a growing corpus of scientific research has been built that focus on the attempt to infer
parameters by reconstructing them from statistical observations of systems. The problem itself is known 
as statistical inference and traces back to the times when the mathematical-physics description 
of nature became fully operative thanks to the advances of mechanics and calculus, i.e. with the French 
mathematicians Laplace and Lagrange. In recent times this field and its most ambitious problems have 
deeply connected with statistical physics \cite{HN,MM,H01} at least in those cases in which the structure of
the problem include the assumption of an underlying model to describe the investigated phenomena.
The aforementioned connection is surely related to the ability that statistical physics has acquired to 
describe phase transitions. In this paper we study the inverse problem for a model of interacting monomer-dimers 
in the mean-field, i.e. in the complete, graph. The denomination comes from the fact that the standard calculation
in statistical mechanics, i.e. the derivation of the free energy and correlation from the assignment of the parameters
is called the direct problem. Monomer-dimer models appeared in equilibrium statistical mechanics
to describe the process of absorption of monoatomic or diatomic molecules in condensed matter
lattices \cite{JKR38}. 
From the physical point of view monomers and dimers cannot occupy the same site
of the lattice due to the hard-core interaction i.e. the strong contact repulsion generated
by the Pauli exclusion principle. Beside such interaction though, as first noticed by Peierls \cite{P36}, 
the attractive component of the Van der Waals potentials might influence the phase
structure of the model and the thermodynamic behaviour of the material. In the mean field setting analysed here the monomer-dimer model displays the phenomenon
of phase coexistence among the two types of particles \cite{ACM14,ACFM14,ACM14bis}. This makes the inverse problem 
particularly challenging since in the presence of phase coexistence the non uniqueness of its solution
requires a special attention in identifying the right set of configurations. Under mean-field theory, the monomer-dimer model can be solved for the monomer densities and the correlations between monomers and dimers: the mean-field solution is inverted to yield the parameters of the model (external field and imitation coefficient) as a function of the empirical observables.
The inverse problem has also been known for a long time as Boltzmann machine learning \cite{AHS85}. Its renewed interest 
is linked to the large number of applications in many different scientific fields like biology \cite{OG91,MC11,RTH09,SM09}, computer science for 
the matching problem \cite{KS81,ZM06,BLS13} and also social sciences \cite{BCSV14,BCFVV15}.

In this paper we follow an approach to the inverse problem similar to the one introduced for the multi-species 
mean-field spin model in the work \cite{MF}. The paper is organised in the following chapters and results.
In the second section we recall briefly the monomer-dimer model and we review the basic properties of its solution \cite{ACM14,ACM14bis}. In the third section we solve the inverse problem: using the monomer density and the susceptibility of the model, we compute the values of the two parameters, here called coupling constants, $J$ and $h$. The first measure the preference for a 
vertex to be occupied by a monomer (respectively dimer), by imitating his neighbours. Firstly we identify the analytical inverse formulas providing an explicit expression of the free parameters in terms of the mentioned macroscopic thermodynamic variables. Then we use the maximum likelihood estimation procedure in order to provide an evaluation of the macroscopic variables starting from real data. The fourth section presents and discusses a set of numerical tests for finite number of particles and finite
number of samples. The dependence of the monomer density and
the susceptibility is studied with respect to the system size. We
find that both of them have a monotonic behavior which depends
on the parameters value and reach their limiting values with a
correction that vanishes as at the inverse volume. We then investigate how the experimental monomer density and susceptibility at fixed volume depend on the number of samples. The effectiveness of the inversion is tested for different values of the imitation coefficients and external fields. After observing that the error of the inversion does not vanish when the parameters are close
to the coexistence phase we investigate the effectiveness of clustering algorithms to overcome the difficulty.  We find
in all cases that the inverse method reconstructs, with a modest amount of samples, the
values of the parameters with a precision of a few percentages. The paper has two technical appendices: the first on the 
rigorous derivation of the exact inverse formulas, the second that supports the first and studies the non homogeneous Laplace 
method convergence to the second order. 
\section{Definition of the model}
Let $G=(V,E)$ be a finite simple graph with vertex set $V$ and edge set $E=\{uv\equiv \{u,v\} | u\neq v\in V\}$.\\
\begin{defn} A \emph{dimer configuration D} on the graph $G$ is a set of \emph{dimers} (pairwise non-incident edges):
$$D \subseteq E\quad \text{and}\quad (uv\in D\Rightarrow uw\notin D\quad\forall w\neq v).$$
The associated set of \emph{monomers} (dimer-free vertices), is denoted by
$$\mathcal{M}(D):=\mathcal{M}_G(D):=\{u\in V|uv\notin D, \forall v\in V\}.$$
\end{defn}
Given a dimer configuration $D\in \mathcal{D}_G$, we set for all $v\in V$ and $e\in E$
\begin{align*}
\alpha_v(D):=
 \begin{cases}
 1,\quad & \text{if}\quad v\in\mathcal{M}(D)\\
 0,\quad &\text{otherwise}
  \end{cases}
\end{align*}
and
\begin{align*}
\alpha_e(D):=
 \begin{cases}
 1,\quad & \text{if}\quad e\in D\\
 0,\quad & \text{otherwise}.
  \end{cases}
\end{align*}
\begin{defn} Let $\mathcal{D}_G$ be the set of all possible dimer configurations on the graph $G$. The \emph{imitative monomer-dimer model} on $G$ is obtained by assigning an external field $h\in\mathbb{R}$ and an imitation coefficient $J\geq 0$ which gives an attractive interaction among particles occupying neighbouring sites. The \emph{Hamiltonian} of the model is defined by the function $H_{G}^{\textsc{imd}}:\mathcal{D}_G\rightarrow\mathbb{R}$ such that 
\begin{equation}\label{eq1_1}
 H_{G}^{\textsc{imd}}:=-\sum_{v\in V}h\alpha_v-\sum_{uv\in E}J(\alpha_u\alpha_v+(1-\alpha_u)(1-\alpha_v)).
\end{equation}
The choice of the Hamiltonian naturally induces a Gibbs probability measure on the space of configuration $\mathcal{D}_G$:
\begin{equation}\label{eq1_2}
 \mu_G^{\textsc{imd}}(D):=\dfrac{\exp(-H_G^{\textsc{imd}}(D))}{Z_G^{\textsc{imd}}}\quad \forall D\in\mathcal{D}_G,
\end{equation}
where the \emph{partition function}$$Z_G^{\textsc{imd}}=\sum_{D\in\mathcal{D}_G}\exp(-H_G^{\textsc{imd}}(D))$$ is the normalizing factor.\\
The natural logarithm of the partition function is called \emph{pressure function} and it is related to the free energy of the model. 
\end{defn}
The normalized expected fraction of monomers on the graph is called \emph{monomer density}. It can also be obtained computing the derivative of the pressure per particle with respect to $h$:
$$m_G^{\textsc{imd}}:=\sum_{D\in\mathcal{D}_G}\dfrac{|\mathcal{M}(D)|}{|V|}\mu_G^{\textsc{imd}}(D)=\dfrac{\partial}{\partial h}\dfrac{\log Z_G^{\textsc{imd}}}{|V|}.$$
It is easy to check that
\begin{equation}\label{eq1_3}
2|D|+|\mathcal{M}(D)|=|V|.
\end{equation}
In this paper we study the imitative monomer-dimer model on the complete graph, that is
$$G=K_N=(V_N,E_N)$$
with $V_N =\{1,\ldots ,N\}$ and $E_N=\{\{u,v\}|u,v\in V_N,u<v\}$.\\
In order to keep the pressure function of order $N$, it is necessary to normalize the imitation coefficient by $\frac{1}{N}$ because the number of edges grows like $N^2$ and to subtract the term $\log N \sum_{e\in E_N}\alpha_e$ to the external field. Thus we will consider the Hamiltonian $H_N^{\textsc{imd}}:\mathcal{D}_N\rightarrow\mathbb{R},$
\begin{equation}
H_N^{\textsc{imd}}:=-\sum_{v\in V_N}h\alpha_v+\log N\sum_{e\in E_N}\alpha_e-\sum_{uv\in E_N}\dfrac{J}{N}(\alpha_u\alpha_v+(1-\alpha_u)(1-\alpha_v)) \; .
\end{equation}
All the thermodynamic quantities will therefore be functions of $N$ and we are interested in studying the large volume limits.\\\\
Before studying the inverse problem, we briefly recall the main properties of the model (see \cite{ACM14,ACM14bis}).\\
Taking $m\in[0,1]$, the following variational principle holds$$p^{\textsc{imd}}=\sup_m \tilde{p}(m),$$
where $p^{\textsc{imd}}$ is the pressure of the model at the thermodynamic limit and $$\tilde{p}(m(J,h),J,h):=-Jm^2+\dfrac{1}{2}J+p^{\textsc{md}}((2m-1)J+h)\quad\forall m\in\mathbb{R},$$
with $p^{\textsc{md}}(\xi):=-\dfrac{1-g(\xi)}{2}-\dfrac{1}{2}\log(1-g(\xi))= -\dfrac{1-g(\xi)}{2}-\log(g(\xi))+\xi\quad\forall\xi\in\mathbb{R}$ and $g(\xi):=\dfrac{1}{2}(\sqrt{e^{4\xi}+4e^{2\xi}}-e^{2\xi})\quad\forall\xi\in\mathbb{R}$.
The solution of the model reduces to identify the value $m^*$ that maximizes the function $\tilde{p}$ and it is found among the solutions of the consistency equation $m=g((2m-1)J+h)$ that include, beside the equilibrium value, also the unstable
and metastable points. It is possible to prove that $m^*$ (which represents the monomer density) is a smooth function for all the values of $J$ and $h$ with the exception of the coexistence curve $\Gamma(J,h)$. Such curve is differentiable in the half-plane
$(J, h)$ which stems from the critical point $(J_c,h_c)=(\frac{1}{4(3-2\sqrt{2})},\frac{1}{2}\ln(2\sqrt{2}-2)-\frac{1}{4}).$

\section{The inverse problem}
The evaluation of the parameters of the model starting from real data is usually called inverse problem and amounts of two steps. The analytical part of the inverse problem is the computation of the values $J$ and $h$ starting from those of the first and second moment of the monomer (or dimer) density. The statistical part instead is the estimation of the values of the moments starting from the real data
and using the maximum likelihood principle \cite{F25} or the equivalent formulations in statistical mechanics terms \cite{J57}. For what it concerns the analytical part, using the results of Appendix A and B, it can be proved that in the thermodynamic limit the imitation coefficient and the external field can be respectively computed as
\begin{equation}\label{eq2_3}
J=-\dfrac{1}{2\chi}+\dfrac{2-m^*}{4m^*(1-m^*)} ,
\end{equation}
and
\begin{equation}\label{eq2_4}
h=g^{-1}(m^*)-J(2m^*-1)=\dfrac{1}{2}\log\left(\dfrac{m^{*2}}{1-m^*}\right)-J(2m^*-1).
\end{equation}
We denote by $m_N$ and $\chi_N$ the finite size monomer 
density average and susceptibility $N(\langle m_N^2\rangle-\langle m_N\rangle^2)$, while their limiting values are denoted without the subscript $N$. 

For the statistical part we use the maximum likelihood estimation procedure. Given a sample of $M$ independent dimer configurations $D^{(1)},\ldots,D^{(M)}$ all distributed according to the measure of Gibbs $\eqref{eq1_2}$, the maximum likelihood function  is defined by
\begin{align*}L(J,h)=\mu_N^{\textsc{imd}}\{ D^{(1)},\ldots,D^{(M)}\}=\prod_{i=1}^M\dfrac{\exp(-H_N^{\textsc{imd}}(D^{(i)}))}{\sum_{D\in\mathcal{D}_{K_N}}\exp(-H_N^{\textsc{imd}}(D))}.
\end{align*}
The function $L(J,h)$ reaches its maximum when the first and the second momentum of the monomer density are calculated from the data according to the following equations:
\begin{align}\label{eq2_7}
\begin{cases}
m_N&=\dfrac{1}{M}\sum_{i=1}^M m_N(D^{(i)}) , \\
m_N^2&=\dfrac{1}{M}\sum_{i=1}^M m_N^2(D^{(i)}).
\end{cases}
\end{align}
The inverse problem is therefore solved by the composition of $\eqref{eq2_7}$ with $\eqref{eq2_3}$ and $\eqref{eq2_4}$. In particular, denoting by $m_{exp}$ and $\chi_{exp}$ respectively the average monomer density and the susceptibility computed from the sample
\begin{equation}\label{eq2_7bis}
m_{exp}=\dfrac{1}{M}\sum_{i=1}^M m_N(D^{(i)})\quad\text{and}\quad \chi_{exp}=N\left(\dfrac{1}{M}\sum_{i=1}^M m_N^2(D^{(i)})-m^2_{exp}\right),\end{equation}
the estimators of the model's free parameters are 
\begin{equation}\label{eq2_8}
J_{exp}=-\dfrac{1}{2\chi_{exp}}+\dfrac{2-m_{exp}}{4m_{exp}(1-m_{exp})}
\end{equation}
and
\begin{equation}\label{eq2_9}
h_{exp}=\dfrac{1}{2}\log\left(\dfrac{m^2_{exp}}{1-m_{exp}}\right)-J_{exp}(2m_{exp}-1).
\end{equation}

%

\section{The inversion at finite volume and finite sample size}
The aim of this chapter is to study the robustness of the inversion procedure, i.e. the computation of the parameters from real data. The idea is to infer the value of $J$ and $h$ from the configurations generated according to the distribution of the model.
In order to compute efficiently the values of the statistical estimators $m_{exp}$ and $\chi_{exp}$ and in order to obtain a good approximation of the analytical inverse formulas in terms of finite size thermodynamic variables, we have to choose a large number of configurations of the sample and a large number of vertices of the graph, which are respectively identified by $M$ and $N$.
Since in real data we have a finite number of vertices and a finite number of configurations, the robustness will be studied with respect to both these two quantities.\\
The data that we are going to use are extracted from a virtually exact simulation of the equilibrium distribution. 
In fact, the mean-field nature of the model allows to rewrite the Hamiltonian $\eqref{eq1_1}$ as a function of the dimer, or monomer, density (see $\eqref{eq1_3}$):
\begin{equation}
H_N^{\textsc{imd}}(d_N)=-N\left(J\left(16d_N^2-4d_N+\dfrac{N-1}{2N}\right)+h\left(1-4d_N\right)-2d_N\log N\right),
\end{equation}
where $d_N=d_N(D)=\frac{|D|}{2N}$, or equivalently
\begin{equation}
H_N^{\textsc{imd}}(m_N)=-N\left(J\left(m_N^2-m_N+\frac{N-1}{2N}\right)+hm_N+\frac{1}{2}\log N(m_N-1)\right).
\end{equation}

\begin{figure}
\centering
\includegraphics[scale=0.5]{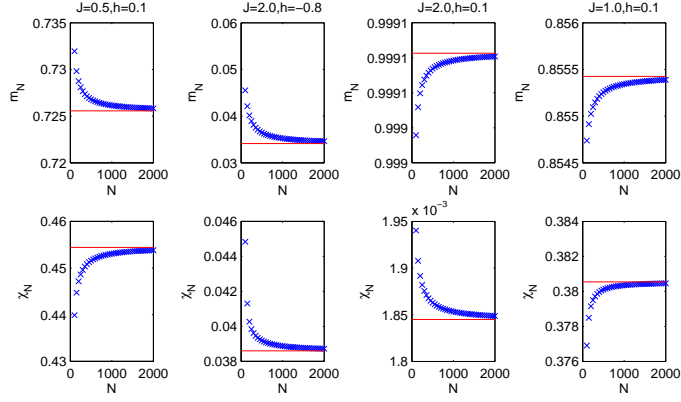}
\caption{Finite size average monomer density $m_N$ (upper panels) and susceptibility $\chi_N$ (lower panels) as a function of $N$ for the monomer-dimer model at different values of $J$ and $h$. The red continuous lines represent the values in the thermodynamic limit.\label{Figure1}}
\end{figure}

\begin{figure}
\centering
\includegraphics[scale=0.7]{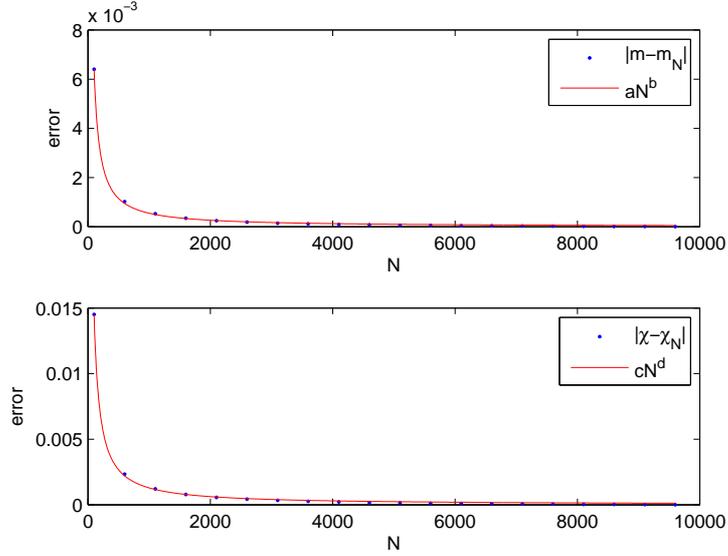}
%
%
%
\caption{$J=0.5,h=0.1.$ Upper panel: $|m_N-m|$ as a function of $N$ together with the
best fit $aN^{b}$ for the data in the left upper panel of figure \ref{Figure1}. We obtain $a = 0.306,a\in(0.1703, 0.4418)$ and $b=-0.8549,b\in(-0.9459, -0.7639)$ with a goodness of fit $R^2 = 0.9815$. Lower panel: $|\chi_N- \chi|$ as a
function of $N$ together with the best fit $cN^d$ for the data in the left lower panel of figure \ref{Figure1}.
We obtain $c = 1.277,c\in(0.9883, 1.566)$ and $d = -0.9765,d\in (-1.024, -0.929)$ with a goodness of fit $R^2 = 0.9971$.
\label{Figure2}}
\end{figure}

\begin{figure}
\centering
\includegraphics[scale=0.7]{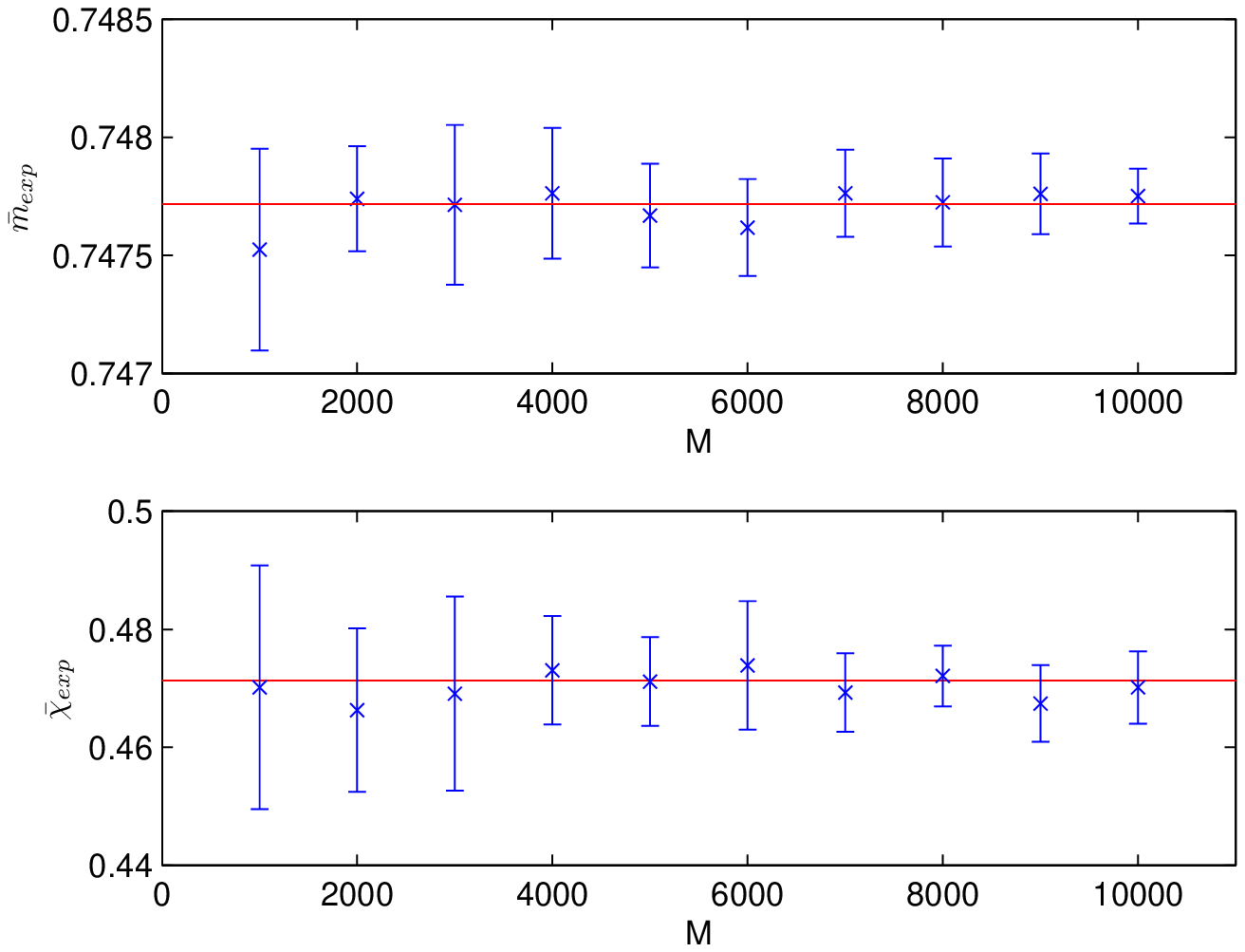}
\caption{$N = 2000$, $J = 0.6$ and $h = 0.1$. Error bars are standard deviations on 20 different $M-$samples of the same simulation. Upper panel: average monomer density
$\overline{m}_{exp}$ (blue crosses) as a function of $M$ (number of the configurations in the sample). The
red continuous line represents the finite size monomer density $m_N$. Lower
panel: susceptibility $\overline{\chi}_{exp}$ (blue crosses) as a function of $M$ (number of the configurations
in the sample. The red continuous
line represents the finite size susceptibility $\chi_N$.\label{Figure3}}
\end{figure}

\begin{figure}
\centering
\includegraphics[scale=0.55]{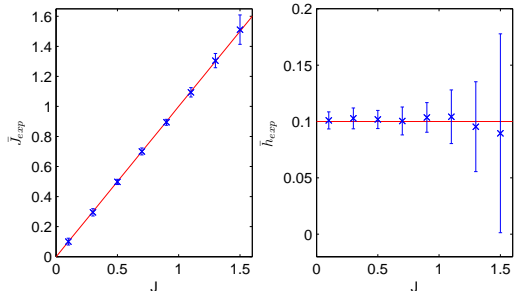}
\caption{Error bars are standard deviations on 20 different $M-$samples of the same simulation. Left panel: $\overline{J}_{exp}$ as a function of $J\in[0.1,1.5]$ (blue crosses). The red continuous line corresponds to the exact value of the imitation coefficient. Right panel: the value of $\overline{h}_{exp}$ (blue crosses) calculated from $\eqref{eq2_9}$ for the values of $J_{exp}$ in the left panel, as a function of $J\in[0.1,1.5]$. The red continuous line corresponds to the exact value of $h$.\label{Figure4}}
\end{figure}

\begin{figure}
\centering
\includegraphics[scale=0.5]{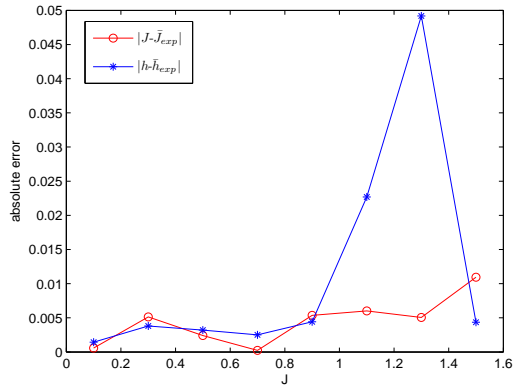}
\caption{Inference of parameters of the monomer-dimer model on 20 different $M-$samples of the same simulation. Absolute errors in reconstructing $J$ and $h$, where $J\in[0.1,1.5]$ and $h=0.1$.\label{Figure5}}
\end{figure}

In particular we use the following definition of the partition function:
\begin{equation}
Z_N^{\textsc{imd}}=\sum_{|D|=0}^{[N/2]}c_N(D)e^{-H_N^{\textsc{imd}}(d_N(D))},
\end{equation}
where the term $c_N(D)=\frac{N!}{|D|!(N-2|D|)!}2^{-|D|}$ is the number of the possible configurations with $|D|$ dimers on the complete graph with $N$ vertices. Using the previous representation of the partition function we extract large samples of dimer densities values according to the equilibrium distribution. Those will be used for the statistical estimation of the first two moments 
$\eqref{eq2_7}$.
We are going to illustrate the results with some examples.
Figure $\ref{Figure1}$ shows the finite size average monomer density $m_N$ and finite size susceptibility $\chi_N$ for the monomer-dimer model at different $N$’s for different couples of parameters $(J,h)$. The figure highlights the monotonic behavior of $m_N$ and $\chi_N$ as function of $N$. We point out that the different monotonic behaviors of the finite size monomer density and susceptibility provide a useful information about the phase space region at which the system is found before applying the full inversion procedure.
\noindent Figure \ref{Figure2} shows the power-law fits of the behavior of the finite size corrections both for monomer density and susceptibility.
In order to test numerically our procedure, we consider $20$ $M-$samples for each couple $(J, h)$ and we solve the inverse problem for each one of them independently; then we average the inferred values over the $20$ $M-$samples. We denote by $\overline{m}_{exp}$, $\overline{\chi}_{exp}$, $\overline{J}_{exp}$ and $\overline{h}_{exp}$ such averaged quantities. The two panels of figure \ref{Figure3} represent the statistical dependence of the estimators $\overline{m}_{exp}$ and $\overline{\chi}_{exp}$ on the number of the configurations of the sample. To check out that dependence on the sample $D^{(1)},\ldots,D^{(M)}$, we computed the values of the experimental estimators over a set of $20$ independent instances of such samples. The errors are standard deviations on 20 different $M-$samples of the same simulation: we find numerical evidence that $M \geq 5000$ stabilizes the estimations.
\\
To test numerically the inversion procedure, we take a sample of $M = 5000$ dimer configurations $\{D^{(i)}\}$, $i=1,\ldots,M$ over a complete graph with $N = 2000$ vertices.
We consider $J\in[0.1,1.5]$ and we fix $h = 0.1$; the obtained values for this case are shown in the left panel of figure \ref{Figure4}, where $\overline{J}_{exp}$ and $\overline{h}_{exp}$ are plotted as functions of $J$. Note that the inferred values of the parameters are in optimal agreement with the exact values. Observe that for large values of $J$, the reconstruction get worse since the interaction between particles grows.\\
In figure \ref{Figure5} we represent the absolute errors as a function of the imitation coefficient in reconstructing $J$ and $h$ in the cases of figure \ref{Figure4}.

\begin{figure}
\centering
\subfigure[$N=100$]
{\includegraphics[width=5cm]{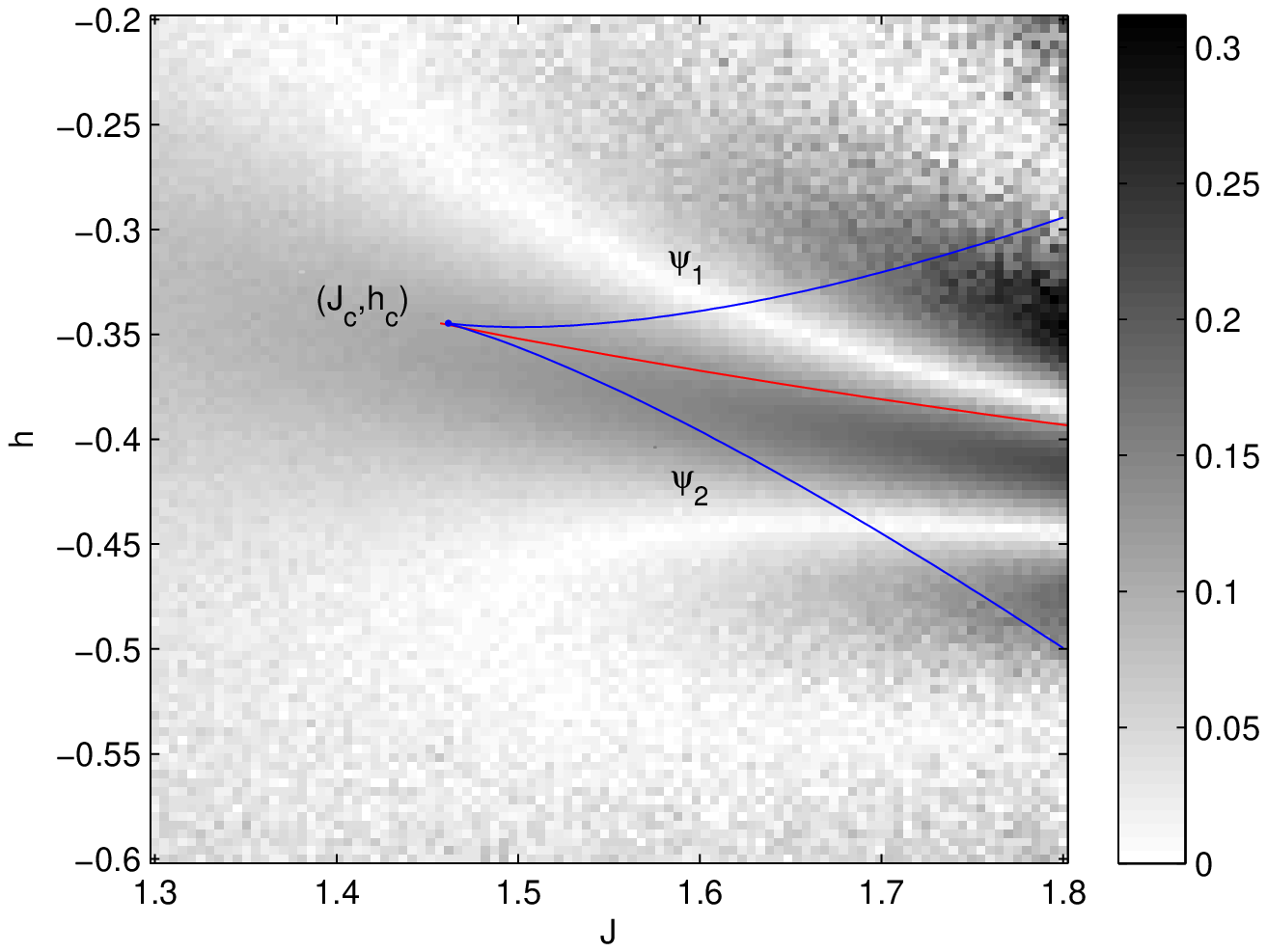}}
\hspace{1mm}
\subfigure[$N=100$]
{\includegraphics[width=5cm]{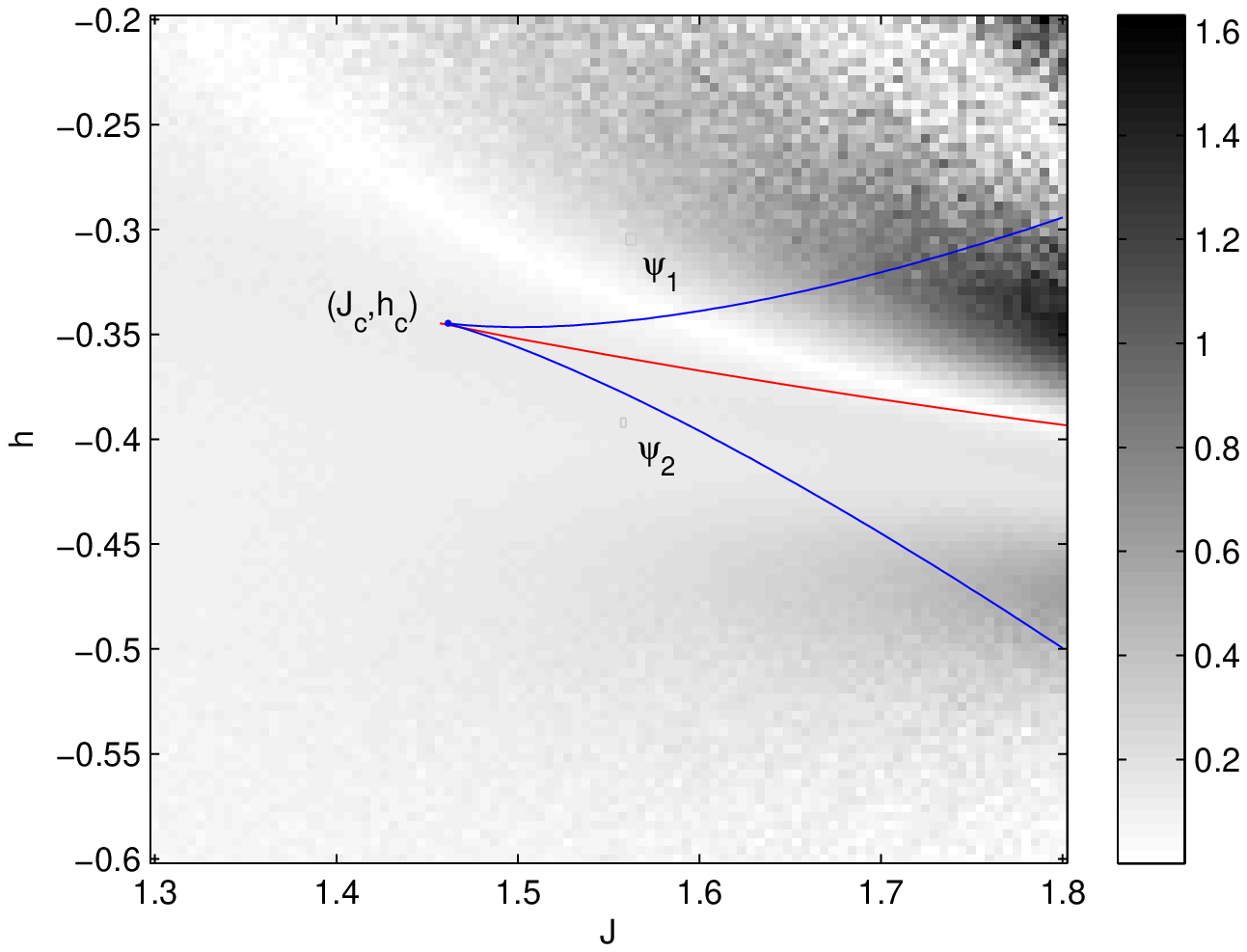}}
\hspace{1mm}
\subfigure[$N=500$]
{\includegraphics[width=5cm]{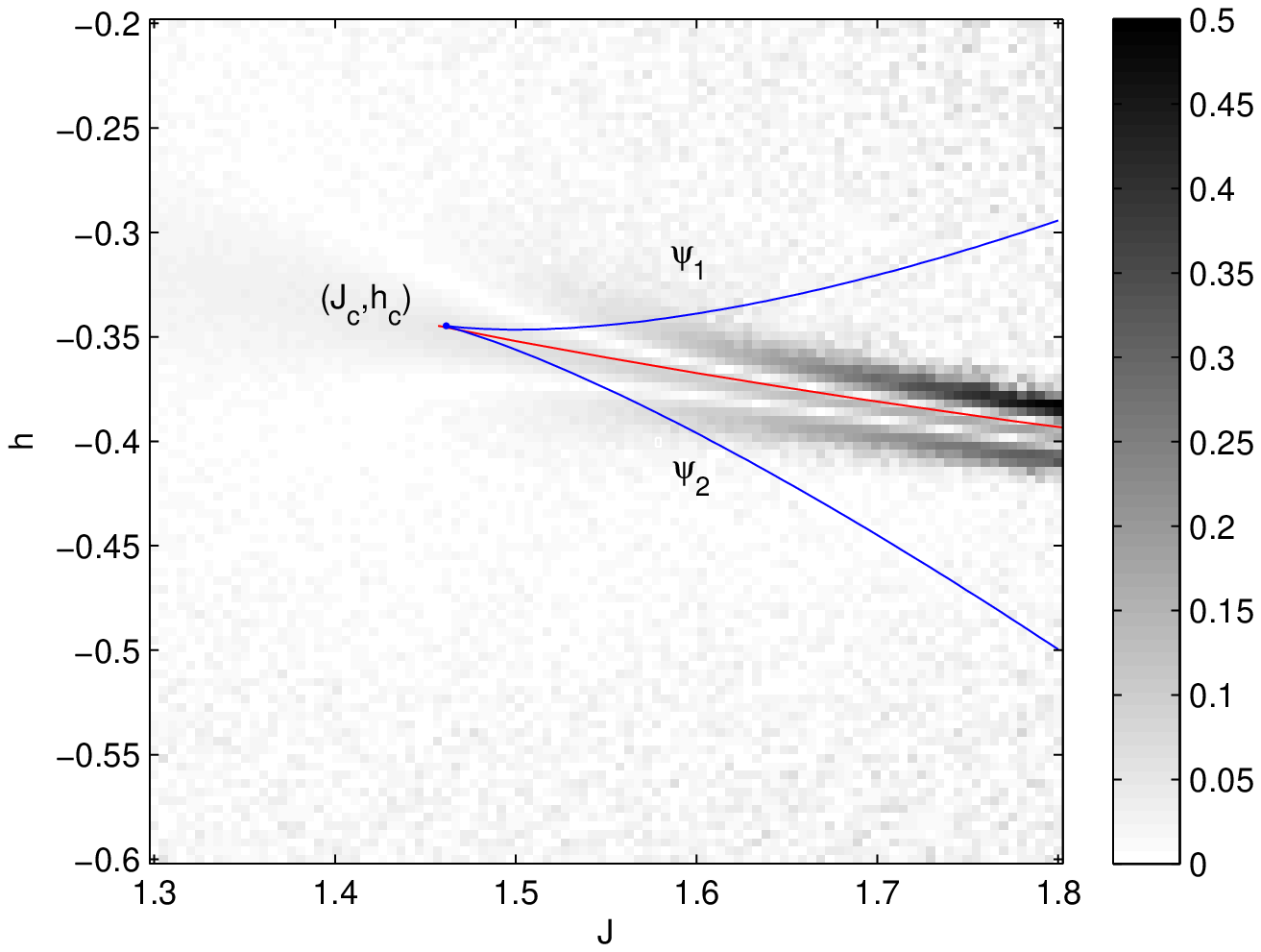}}
\hspace{1mm}
\subfigure[$N=500$]
{\includegraphics[width=5cm]{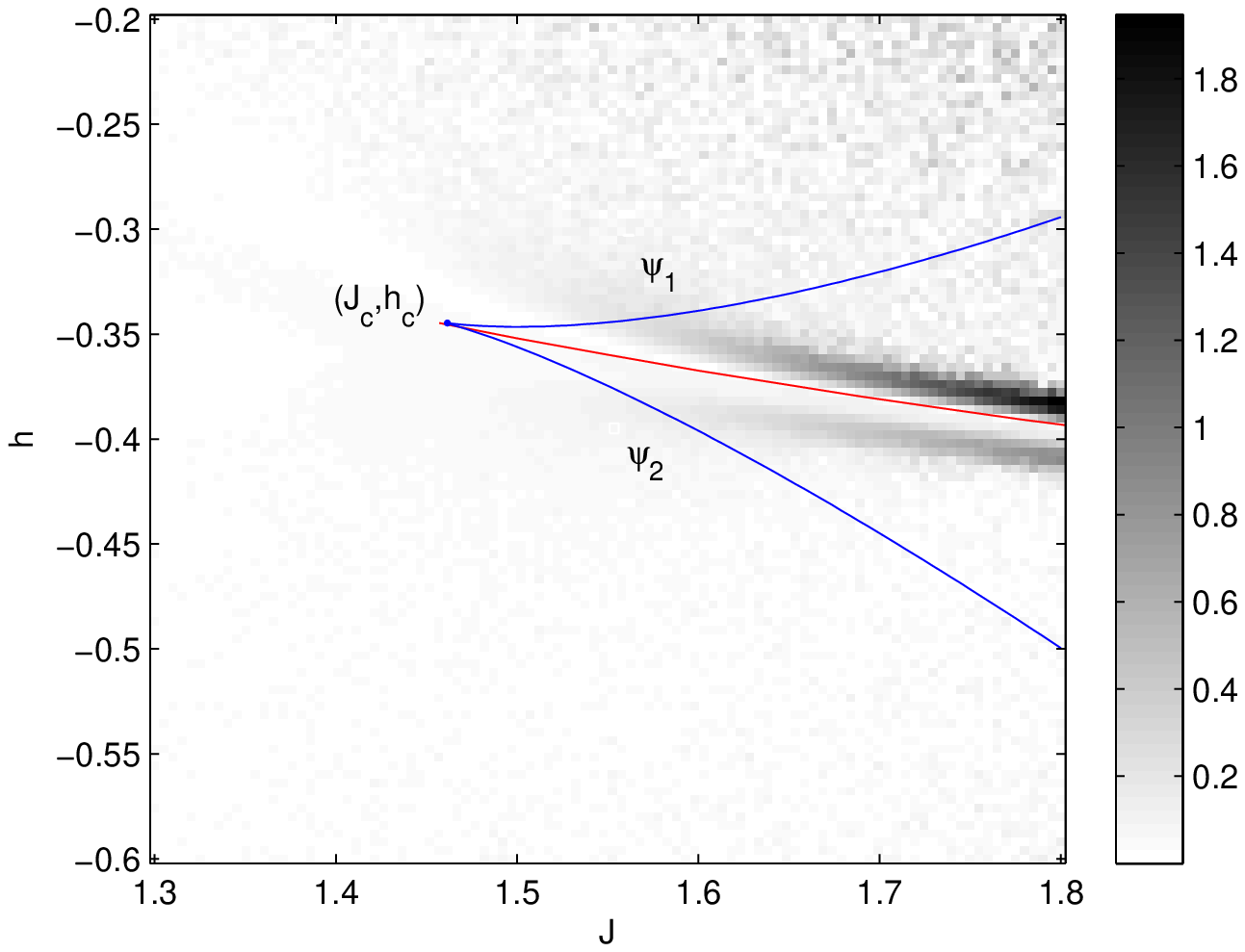}}
\hspace{1mm}
\subfigure[$N=1000$]
{\includegraphics[width=5cm]{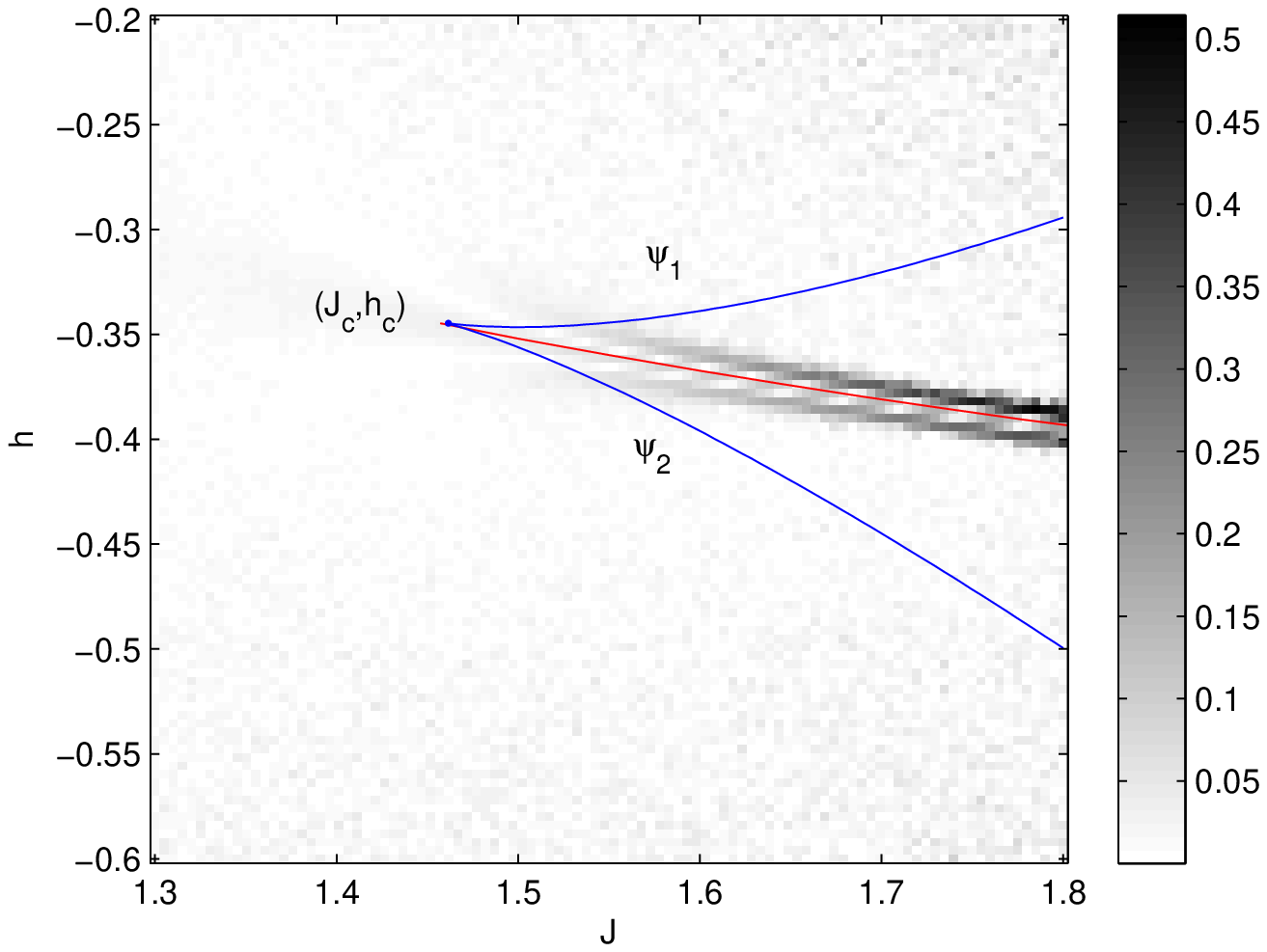}}
\hspace{1mm}
\subfigure[$N=1000$]
{\includegraphics[width=5cm]{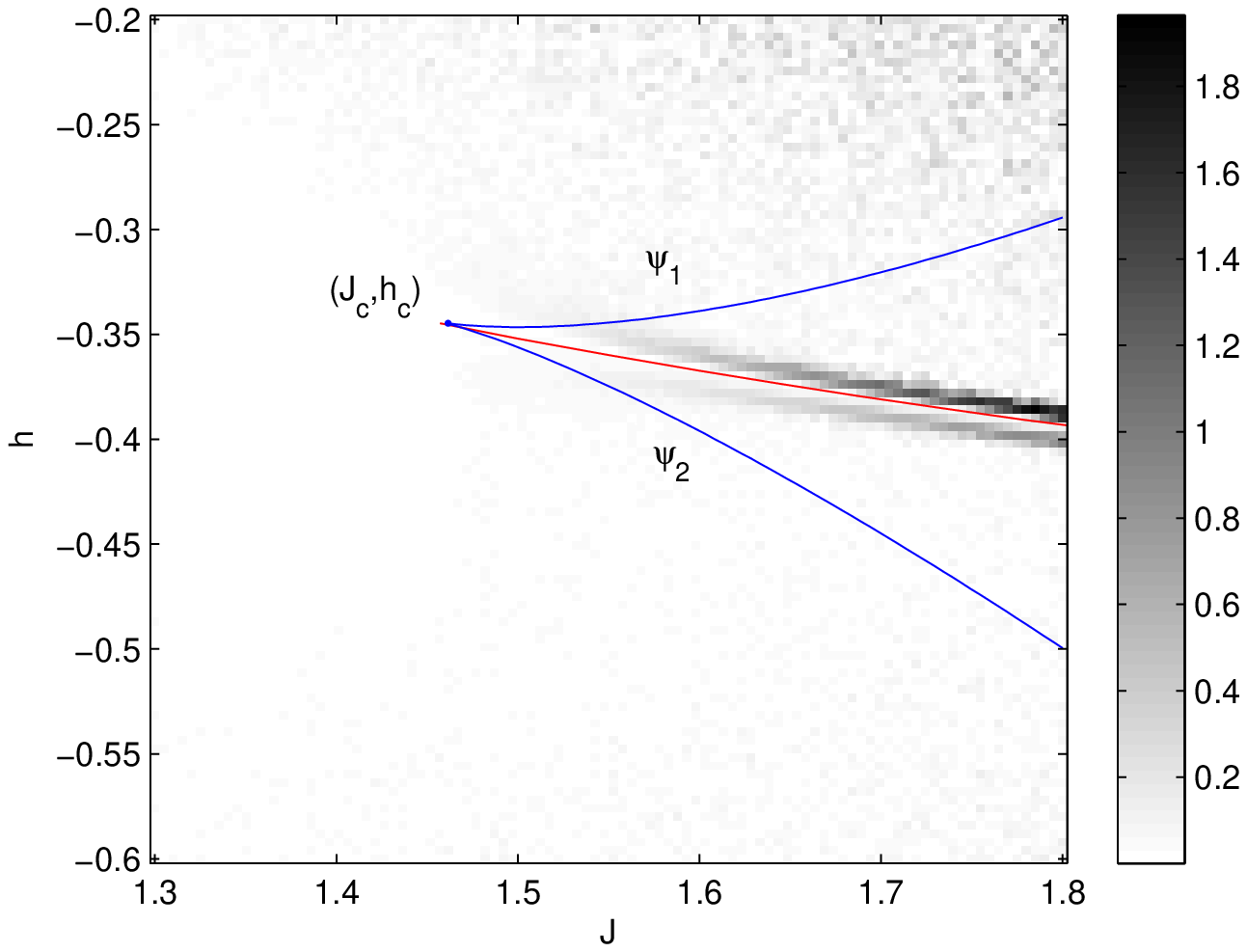}}
\hspace{1mm}
\subfigure[$N=3000$]
{\includegraphics[width=5cm]{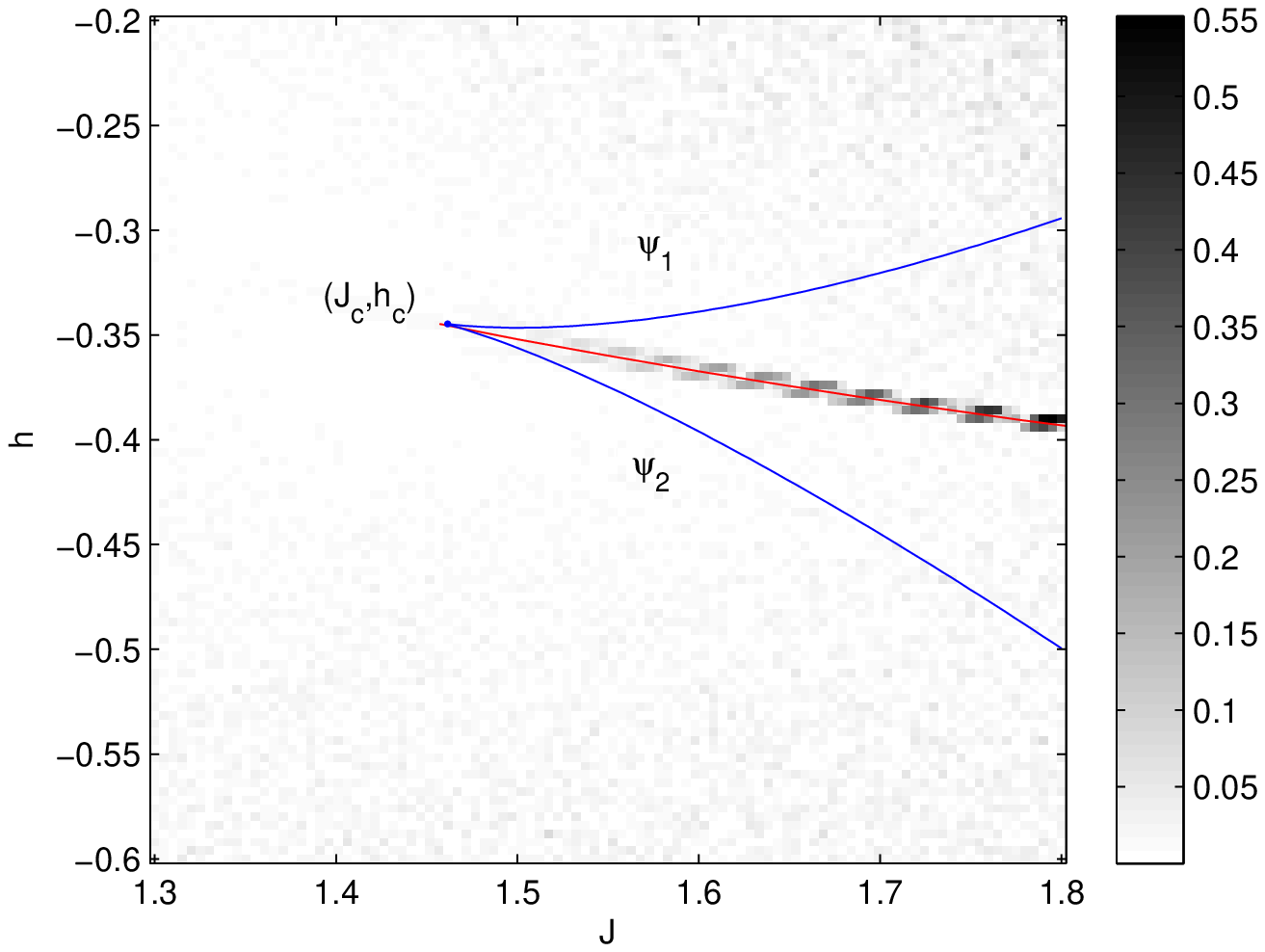}}
\hspace{1mm}
\subfigure[$N=3000$]
{\includegraphics[width=5cm]{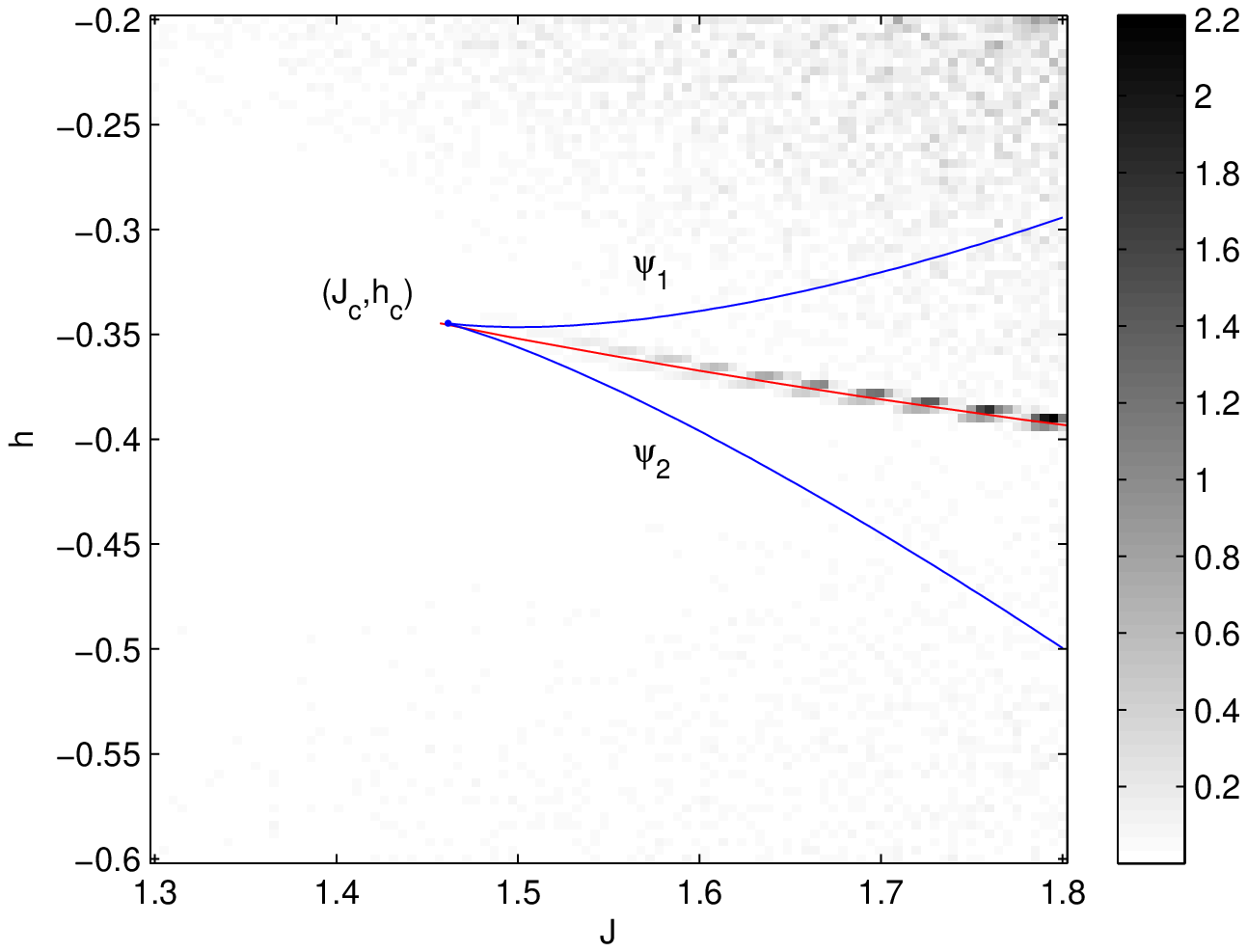}}
\caption{Left panels: relative errors in reconstructing the imitation coefficient $J$. Right panels: relative errors in reconstructing the external field $h$. The points of the phase space are coloured with respect to the errors which assume the highest values along the coexistence line. The graybar on the right gives a range for the computed errors: the scale goes from white for the lowest to black for the highest. The blue curves $\psi_1$ and $\psi_2$ define the region of metastability (see \cite{ACM14,ACFM14,ACM14bis}), the red curve is the coexistence line while $(J_c,h_c)$ is the critical point; the blue and red colors do not identify any error. The number of configurations of the sample is set to be $M=500$. \label{Figure6}}
\end{figure}

Figure \ref{Figure6} shows relative errors in recostructing parameters for increasing sizes of the graph. It highlights that for large values of $N$ and $M$, the inference of parameters doesn't give good results only in the case that the couple $(J,h)$ is close to the coexistence line, but when we deal with real data, it may happen that we don't have a model defined over a graph with a large number of vertices or numerous configurations of the sample. In these cases, when $J$ and $h$ take values in the region of metastability, the inversion at finite volume and finite sample size can't be made using the method descripted above and we need another procedure to solve the problem, as it is shown in the following section.

\section{The inversion at finite volume and finite sample size with clustered phase space}
We are now going to work over the monomer-dimer inverse problem when the phase space doesn't present only one equilibrium state, i.e. when the system undergoes a phase transition. We explain how to modify the mean-field approach we have seen above.
If the model is defined for the parameters $J$ and $h$ such that the couple $(J,h)\in\Gamma$, the Gibbs probability density of the model presents two local maxima and we cannot study the inversion problem in a global way as we have done in the second section but we have to understand what happens in a local neighborhood of each maximum. 
 Given $M$ independent dimer configurations $D^{(1)},\ldots,D^{(M)}$ all distributed according to the Gibbs probability measure for this model,  we can understand their behavior around $m_1$ and $m_2$ separating them in two sets, before applying formulas $\eqref{eq2_8}$ and $\eqref{eq2_9}$, i.e. we divide the configurations of the sample in clusters using the so called clustering algorithms which classify elements into classes with respect to their similarity (see \cite{RL14,MKD03,DR16,NB12}). The clustering algorithms we use are based on the distance between the monomer density of the configurations: we put them in the same group if they are close enough and far from the other clusters (the concept of distance between clusters will be discussed later).\\
The method we use is the density clustering \cite{RL14}, which is based on the idea that the cluster centers are encircled by near configurations with a lower local density and that they are relatively far from any other configuration with a high local density. For each configuration we compute two quantities: its local density $\rho_i$ and its distance $\delta_i$ from configurations with higher density. These quantities depend on the euclidean distance $d_{ij}=|m^{(i)}-m^{(j)}|$, where $m^{(i)},$ for $i=1,\ldots,M$ is the monomer density of the configuration $D^{(i)}$.\\  
The local density $\rho_i$ of $D^{(i)}$ is defined by
\begin{equation}\label{eq4_1a}
\rho_i=\sum_{j=1}^M \varphi(d_{ij}-d_c),
\end{equation}
where $d_c$ is an arbitrary cutoff distance (we will discuss later the choice of $d_c$) and
\begin{align*}
\varphi(x)=
\begin{cases}
1\qquad &\text{if}\quad x<0\\
0\qquad &\text{otherwise}.\\
\end{cases}
\end{align*}
In other words, the local density $\rho_i$ corresponds to the number of configurations that are closer than $d_c$ to the configuration $D^{(i)}$.
\begin{rem}
The choice of the cutoff distance $d_c$ is crucial for the results of the algorithm: if we take a too large or a too small value for $d_c$ it is possible that the algorithm is not able to find correctly the cluster centers. From the results of our simulations it emerges that, if we want to solve the inverse problem over a complete graph with $N=3000$ vertices working with a sample made of $M=10000$ dimer configurations, we have the best reconstruction of the free parameters when $d_c$ is setted to be equal to $0.01$. Obviously the choice depends on the range where the clusters centers have to be found and on the number of configurations of which the sample is made. More in general we have seen that for large values of $M$, the minimum absolute error in reconstructing parameters occurs when the cutoff distance is equal to $\frac{C}{M}$.
\end{rem}
The distances $\delta_i$ are the minimum distance between the configuration $D^{(i)}$ and any other configuration with higher local density:
\begin{equation}\label{eq4_1b}
\delta_i=\min\limits_{j:\rho_j>\rho_i} d_{ij},\end{equation}
while for the configuration with the highest local density we take $\delta_{\bar{i}}=\max\limits_{j} d_{ij}$.\\
Observe that the quantity $\delta_i$ is much larger than the typical nearest neighbor distance only for the configurations that are local or global maxima in the density. Thus cluster centers are recognised as configurations for which the $\delta_i$ is anomalously large (this situation will be illustrated in example \ref{ex1} in the following).\\
After the cluster centers have been found, each remaining configuration is assigned to its closest neighbor with higher density.
\begin{rem}
We tested our inversion formulas using two other clustering algorithms, obtaining analogous results, which put a number of data points into $K$ clusters starting from $K$ random values for the centers $x^{(1)},\ldots,x^{(K)}$: the $K$-means clustering algorithm and the soft $K$-means clustering algorithm \cite{MKD03}. However the results we are going to talk about have been obtained using the density clustering algorithm: by using this algorithm we do not have to specify the number of clusters since it finds them by itself. 
\end{rem}
\begin{rem}\label{rem2}
From the results of our simulations, according to the example \ref{ex1} in the following, it emerges that, if the couple of parameters which defines the model is not close enough to the coexistence line, we have a better reconstruction of the parameters applying equations $\eqref{eq2_8}$ and $\eqref{eq2_9}$ to the configurations which belong to the largest cluster.\\
On the other hand, when the couple $(J,h)$ is near to the coexistence line $\Gamma(J,h)$, we solve the problem applying equations $\eqref{eq2_8}$ and $\eqref{eq2_9}$ to each cluster and averaging the inferred values as follows. We define the respective observables of the two classes as
$$m_{exp}^{(k)}=\dfrac{1}{M_k}\sum_{i\in \mathcal{C}_k}m_i$$ and $$\chi_{exp}^{(k)}=N\left(\dfrac{1}{M_k}\sum_{i\in \mathcal{C}_k}m_i^2-(m_{exp}^{(k)})^2\right),$$where $k\in\{1,2\}$, $\mathcal{C}_k$ is the set of indices of the configurations belonging to the $k^{th}$ cluster and $M_k=|\mathcal{C}_k|$ is its cardinality.
\\We now apply $\eqref{eq2_8}$ separately to each group in order to obtain two different estimators $J_{exp}^{(1)}$ and $J_{exp}^{(2)}$; finally we take the weighted average of all the different estimates
\begin{equation}\label{eq4_2}
J_{exp}=\dfrac{1}{M_1+M_2}(M_1J_{exp}^{(1)}+M_2J_{exp}^{(2)})
\end{equation}
in order to obtain the estimate for the imitation coefficient.\\
To estimate the parameter $h$, we first compute the values $h_{exp}^{(1)}$ and $h_{exp}^{(2)}$ within each cluster using equation $\eqref{eq2_9}$ and the corresponding $J_{exp}^{(k)}$; the final estimate for $h$ is given by the weighted average over the clusters
\begin{equation}\label{eq4_3}
h_{exp}=\dfrac{1}{M_1+M_2}(M_1h_{exp}^{(1)}+M_2h_{exp}^{(2)}).
\end{equation}
\end{rem}

We now focus on some cases of clustered phase space and we solve the inverse problem applying the density clustering algorithm.\\
In order to test numerically the inversion procedure for the monomer-dimer model, we consider a sample of $M = 10000$ dimer configurations $\{D^{(i)}\}$, $i=1,\ldots,M$ over a complete graph with $N = 3000$ vertices. We denote by the bar averaged quantities and the errors are standard deviations over 20$-M$ samples.

\begin{ex}\label{ex1}
Consider a monomer-dimer model defined by the couple $$(J,h)=(2.001,-0.4145);$$ the Gibbs probability distribution of the monomer densities for this choice of parameters is represented in figure \ref{Figure7}. Given $M=10000$ independent dimer configurations $D^{(1)},\ldots,D^{(M)}$ all distributed according to the Gibbs probability measure for this model, we use the density clustering algorithm in order to divide them in two sets to reconstruct the parameters.\\As we can see by figures \ref{Figure7} and \ref{Figure8}, configurations are divided in two clusters $\mathcal{C}_1$ and $\mathcal{C}_2$ respectively centered in $\overline{m}_1=0.1507\pm5.7\cdot10^{-17}$ and $\overline{m}_2=0.9402\pm9.9\cdot10^{-4}$; moreover the cluster centered in $m_1$ contains more configurations than that centered in $m_2$.
Let start observing that the reconstructed parameters are better solving the problem only respect to the largest cluster.\\
Applying equations $\eqref{eq2_8}$ and $\eqref{eq2_9}$ both to the configurations in $\mathcal{C}_1$ and $\mathcal{C}_2$ according to remark \ref{rem2}, by formulas $\eqref{eq4_2}$ and $\eqref{eq4_3}$ we obtain the following reconstructed values of parameters:
 \begin{equation}\label{eq_4ex1}\overline{J}_{exp}=2.0141\pm0.0802\quad\text{and}\quad \overline{h}_{exp}=-0.4196\pm0.0828.\end{equation}
Applying instead equations $\eqref{eq2_8}$ and $\eqref{eq2_9}$ only to the configurations in the largest cluster $\mathcal{C}_1$, we obtain the following reconstructed values of parameters:
\begin{equation}\label{eq_4ex2}\overline{J}_{exp}=2.0036\pm0.0353\quad\text{and}\quad \overline{h}_{exp}=-0.4091\pm0.0247.\end{equation}

\begin{figure}
\centering
\includegraphics[scale=0.5]{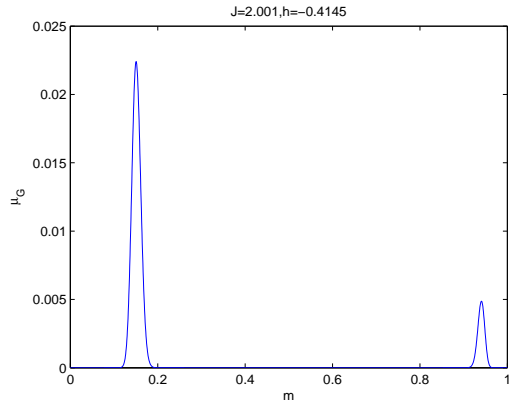}
\caption{Gibbs probability distribution of the monomer densities for the dimer configurations of the monomer-dimer model defined by the couple of parameters $(J,h)=(2.001,-0.4145)$.\label{Figure7}}
\end{figure}
\begin{figure}
\centering
\includegraphics[scale=0.7]{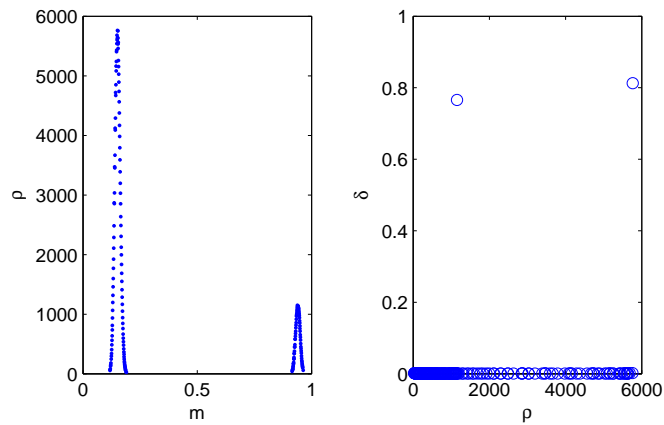}
\caption{Density clustering algorithm. Left panel: plot of the vector $\rho$, whose components are computed according to $\eqref{eq4_1a}$, of the density of configurations around each configurations of the considered sample as a function of the monomer densities. Right panel: decision graph, plot of the vector $\delta$, whose components are computed according to $\eqref{eq4_1b}$, as a function of the vector $\rho$.\label{Figure8}}
\end{figure}
In order to justify our choice for the cutoff distance, we focus on figure \ref{Figure9}, which shows the euclidean distances between $\overline{J}_{exp}$ and the true parameter $J$ (blue stars) and between $\overline{h}_{exp}$ and the true parameter $h$ (red circles) for each choice of $d_c$, that takes value $10^{-j}$, for $j=1,\ldots,6$. We can see that, taking a sample of $M=10000$ dimer configurations over a complete graph with $N=3000$ vertices, we obtain the minimum absolute error considering $d_c=0.01$. According to what we have told above, the choice is arbitrary and it depends on the range of values of the monomer densities and on the number of configurations in the sample: obviously, working with a larger set of dimer configurations we have more freedom in the choice of the cutoff distance.
\begin{figure}[H]
\centering
\includegraphics[scale=0.5]{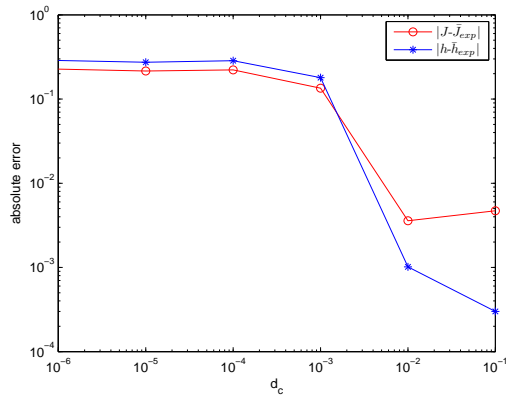}
\caption{Density clustering algorithm: choice of the cutoff distance. Absolute errors in reconstructing $J$ and $h$. Distance between the reconstructed $\overline{J}_{exp}$ and the true value $J$ (blue stars) and distance between $\overline{h}_{exp}$ and $h$ (red circle) for each choice of the cutoff distance $d_c$, which takes value $10^{-j}$, for $j=1,\ldots,6$. The values of $J_{exp}$ and $h_{exp}$ are averaged across $20$ $M-$samples. The errors are plotted as a function of $d_c$.\label{Figure9}}
\end{figure}
In conclusion we have seen that in the case that the couple of parameters $(J,h)$ belongs to the region of metastability and is far enough from the coexistence line, at finite volume and at finite sample size, there are two clusters and one of them is much larger than the other one. According to remark \ref{rem2}, the obtained results confirm that the reconstruction of the parameters is better if we apply formulas $\eqref{eq2_8}$ and $\eqref{eq2_9}$ only to the largest set of configurations. The goodness of results is estimated comparing $\eqref{eq_4ex1}$ and $\eqref{eq_4ex2}$: the distance between the reconstructed parameters $\overline{J}_{exp}$ and the true value $J$ is smaller in the first case, while the respective recontructions of $h$ are equivalent.
\end{ex}
We proceede considering ten different couples of parameters which are nearby the coexistence line $\Gamma(J,h)$ descripted above. In order to define them we take ten equispaced values for the imitation coefficient $J$ in the interval $[1.6,2]$ and we compute the corresponding values for the parameter $J$ using equations $\eqref{eq4_2}$ and $\eqref{eq4_3}$. The obtained values are shown in figure \ref{Figure10}, where  $\overline{J}_{exp}$ and $\overline{h}_{exp}$ are plotted as a function of $J$.
\begin{figure}[H]
\centering
\includegraphics[scale=0.55]{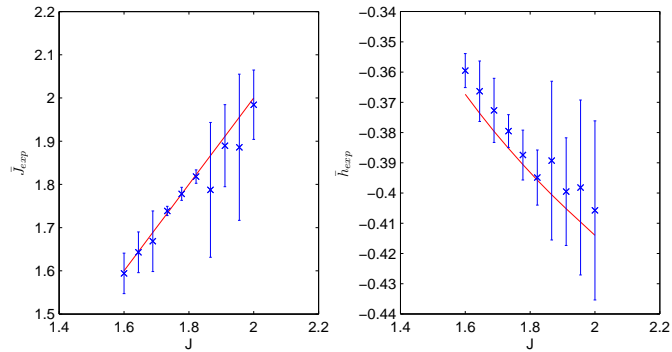}
\caption{$N=3000$, $J\in[1.6,2]$, $h$ takes values over the coexistence line. Error bars are standard deviations on 20 different $M-$samples of the same simulation. Parameters are reconstructed using the density clustering algorithm. Left panel: $\overline{J}_{exp}$ (blue crosses) calculated from $\eqref{eq4_2}$ as a function of $J$.  The red continous line represents the true value of $J$. Right panel: the value of $\overline{h}_{exp}$ (blue crosses) calculated from $\eqref{eq4_3}$ for the values of $\overline{J}_{exp}$ in the left panel, as a function of $J$. The red continuous line corresponds to the exact value of $h$.\label{Figure10}}
\end{figure}
In figure \ref{Figure11} we can see the results in reconstructing parameters crossing the coexistence line $\Gamma(J,h)$. Fixed $J=1.8$ we take increasing values of the parameter $h$ in the interval $[-0.3940,-0.3924]$. In figure \ref{Figure12} we can see how the distribution of Gibbs of the monomer densities changes for different values of $h$.

In figure \ref{Figure13} the euclidean distances between $\overline{J}_{exp}$ and the value $J=1.8$ (blue stars) and between $\overline{h}_{exp}$ and $h\in[-0.3940,-0.3924]$ (red circles) are shown for each of the nine couples $(J,h)$.
\begin{figure}
\centering
\includegraphics[scale=0.55]{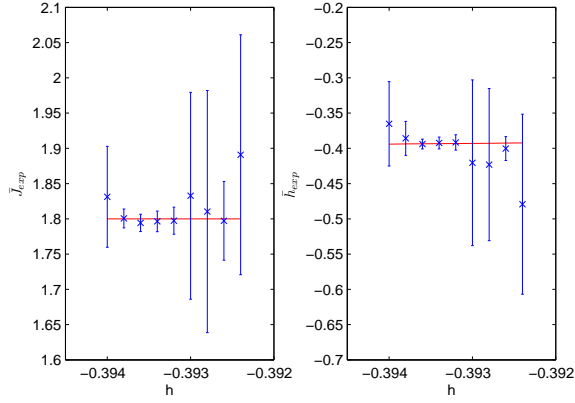}
\caption{$N=3000$, $J=1.8$, $h\in[-0.3940,-0.3924]$. Error bars are standard deviations on 20 different $M-$samples of the same simulation. Parameters are reconstructed using the density clustering algorithm. Left panel: the value of $\overline{J}_{exp}$ (blue crosses) calculated from $\eqref{eq4_2}$ as a function of $h$ together with the statistical error. The red continous line represents the true value of $J$. Right panel: the value of $\overline{h}_{exp}$ (blue crosses) calculated from $\eqref{eq4_3}$ for the values of $\overline{J}_{exp}$ in the left panel, as a function of $h$ together with the statistical error. The red continuous line corresponds to the exact value of $h$.\label{Figure11}}
\end{figure}
\begin{figure}
\centering
\includegraphics[scale=0.7]{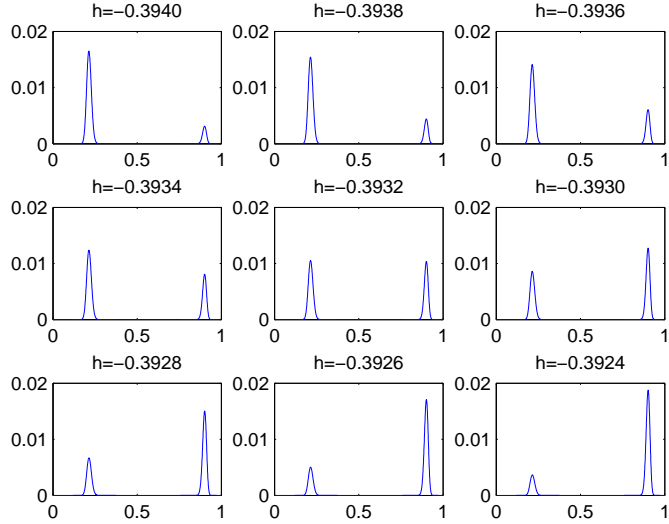}
\caption{$N=3000$, $J=1.8$, $h\in[-0.3940,-0.3924]$. Gibbs probability distribution of the monomer densities for the dimer configurations of the monomer-dimer model defined by each couple of parameters $(J,h)$ defined in figure \ref{Figure11}.\label{Figure12}}
\end{figure}
\begin{figure}
\centering
\includegraphics[scale=0.5]{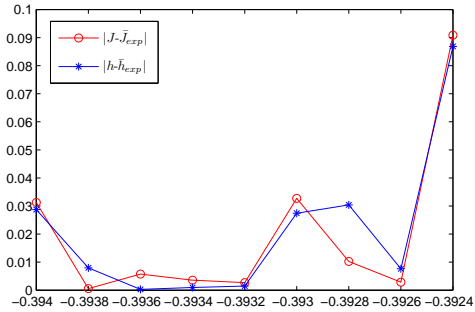}
\caption{$N=3000$, $J=1.8$, $h\in[-0.3940,-0.3924]$. Absolute errors in reconstructing $J$ and $h$ using the density clustering algorithm. Distance between the
reconstructed $\overline{J}_{exp}$ and the true value $J$ (blue stars) and distance between
$\overline{h}_{exp}$ and $h$ (red circle) for each couple of parameters. The values of $\overline{J}_{exp}$ and $\overline{h}_{exp}$ are averaged across $20$ $M-$sample. The errors are plotted as a function of $h$.\label{Figure13}}
\end{figure}

\newpage
\appendix
\section{Monomer-dimer model. Thermodynamic limit of the susceptibility.}
In this appendix, using the extended Laplace's method studied in Appendix B, we prove that 
\begin{equation}\nonumber\lim\limits_{N\rightarrow\infty}\dfrac{\partial}{\partial h}\langle m_N(J,h)\rangle=\dfrac{\partial}{\partial h}m(J,h).\end{equation}
We have used this result in the fourth section.
\begin{thm}
Given an imitative monomer-dimer model defined by a couple of parameters $(J,h)$ over a complete graph of $N$ vertices, it holds:
\begin{equation}\label{eqS1}\lim\limits_{N\rightarrow\infty}\dfrac{\partial}{\partial h}\langle m_N(J,h)\rangle=\dfrac{\partial}{\partial h}m(J,h).\end{equation}
\begin{rem}
According to results in \cite{ACFM14},
write the partition function of the monomer-dimer model as
\begin{equation}
\nonumber Z_N(J,h)=\int_{\mathbb{R}}e^{NF_N(x)}dx,
\end{equation}
where \begin{equation}\label{eqS01}F_N(x)=-Jx^2+p_N^{(0)}((2x-1)J+h),\end{equation}
\begin{equation}\label{eqS02}
p_N(J,h)=\dfrac{1}{N}\log\left(\int_{\mathbb{R}}e^{NF_N(x)}dx\right),
\end{equation}
$$p_N^{(0)}(J,h)=\left.p_N(J,h)\right|_{J=0}.$$
Let $c_N$ be the maximum point of the function $F_N(x)$.
In order to simplify the notations set $\bar{x}:=(2x-1)J+h$ and $\bar{c}:=(1-2c)J+h$.\\
\end{rem}
\proof
Let start computing the expectation of the monomer density using the definition of the pressure function given in \eqref{eqS02}:
\begin{align*}
\langle m_N(J,h)\rangle=&\dfrac{\partial p_N(J,h)}{\partial h}=\dfrac{1}{N}\dfrac{\partial}{\partial h}\log\left(\int_{\mathbb{R}}e^{NF_N(x)}dx\right)=\\
=& \dfrac{\displaystyle\int_{\mathbb{R}}e^{NF_N(x)}\dfrac{\partial}{\partial h}p_N^{(0)}(\bar{x})dx}{\displaystyle\int_{\mathbb{R}}e^{NF_N(x)}dx}.
\end{align*}
The finite size susceptibility can be written as: 
\begin{align}
&\nonumber\chi_N(J,h)=\dfrac{\partial \langle m_N(J,h)\rangle}{\partial h}=\\
\nonumber=& \dfrac{\displaystyle\int_{\mathbb{R}}e^{NF_N(x)}\left[N\left(\dfrac{\partial p_N^{(0)}(\bar{x})}{\partial h}\right)^2+\dfrac{\partial^2 p_N^{(0)}(\bar{x})}{\partial h^2}\right]dx}{\displaystyle\int_{\mathbb{R}}e^{NF_N(x)}dx}-N\dfrac{\left(\displaystyle\int_{\mathbb{R}}e^{NF_N(x)}\dfrac{\partial}{\partial h}p_N^{(0)}(\bar{x})dx\right)^2}{\left(\displaystyle\int_{\mathbb{R}}e^{NF_N(x)}dx\right)^2}=\\
\label{eqS2}=&\dfrac{\displaystyle\int_{\mathbb{R}}e^{NF_N(x)}\dfrac{\partial^2}{\partial h^2}p_N^{(0)}(\bar{x})dx}{\displaystyle\int_{\mathbb{R}}e^{NF_N(x)}dx}+\\
\label{eqS3}&+N\left[\dfrac{\displaystyle\int_{\mathbb{R}}e^{NF_N(x)}\left(\dfrac{\partial}{\partial h}p_N^{(0)}(\bar{x})\right)^2 dx}{\displaystyle\int_{\mathbb{R}}e^{NF_N(x)}dx}-\left(\dfrac{\displaystyle\int_{\mathbb{R}}e^{NF_N(x)}\dfrac{\partial}{\partial h}p_N^{(0)}(\bar{x})dx}{\displaystyle\int_{\mathbb{R}}e^{NF_N(x)}dx}\right)^2\right].
\end{align}
Now we are going to use the extended Laplace's method in order to evaluate the behavior of \eqref{eqS2} and \eqref{eqS3} at the thermodynamic limit.
\\Observe that, since all the quantities computed above are limited, the second order extended Laplace's method suffices to study the behavior of the finite size susceptibility as $N\rightarrow\infty$.
\\
As $N\rightarrow\infty$, the numerator of $\eqref{eqS2}$ can be approximated as:
\begin{align}
\nonumber&\sqrt{\dfrac{2\pi}{-NF''(c)}}e^{NF_N(c_N)}\left\{\left.\dfrac{\partial^2 p^{(0)}(\bar{x})}{\partial h^2}\right|_{\bar{x}=\bar{c}}+\dfrac{1}{N}\left[-\dfrac{\dfrac{\text{d}^2}{\text{d} x^2}\left.\dfrac{\partial^2 p^{(0)}(\bar{x})}{\partial h^2}\right|_{\bar{x}=\bar{c}}}{2F''(c)}+\right.\right.\\
\label{eqS4}&+\left.\left.\dfrac{\left.\dfrac{\partial^2 p^{(0)}(\bar{x})}{\partial h^2}\right|_{\bar{x}=\bar{c}}F^{(iv)}(c)}{8(F''(c))^2}+\dfrac{\dfrac{\text{d}}{\text{d} x}\left.\dfrac{\partial^2 p^{(0)}(\bar{x})}{\partial h^2}\right|_{\bar{x}=\bar{c}}F'''(c)}{2(F''(c))^2}-\dfrac{5\left.\dfrac{\partial^2 p^{(0)}(\bar{x})}{\partial h^2}\right|_{\bar{x}=\bar{c}}(F'''(c))^2}{24(F''(c))^3}\right]\right\}.
\end{align}
As $N\rightarrow\infty$, the numerator of the first fraction in $\eqref{eqS3}$ can be approximated as:
\begin{align}
\nonumber&\sqrt{\dfrac{2\pi}{-NF''(c)}}e^{NF_N(c_N)}\left\{\left.\left(\dfrac{\partial p^{(0)}(\bar{x})}{\partial h}\right)^2\right|_{\bar{x}=\bar{c}}+\dfrac{1}{N}\left[\dfrac{-\dfrac{\text{d}^2}{\text{d} x^2}\left.\left(\dfrac{\partial p^{(0)}(\bar{x})}{\partial h}\right)^2\right|_{\bar{x}=\bar{c}}}{2F''(c)}+\right.\right.
\\
\nonumber&+\dfrac{\left.\left(\dfrac{\partial p^{(0)}(\bar{x})}{\partial h}\right)^2\right|_{\bar{x}=\bar{c}}F^{(iv)}(c)}{8(F''(c))^2}+\dfrac{\dfrac{\text{d}}{\text{d} x}\left.\left(\dfrac{\partial p^{(0)}(\bar{x})}{\partial h}\right)^2\right|_{\bar{x}=\bar{c}}F'''(c)}{2(F''(c))^2}+\\
\nonumber&-\left.\left.\dfrac{5\left.\left(\dfrac{\partial p^{(0)}(\bar{x})}{\partial h}\right)^2\right|_{\bar{x}=\bar{c}}(F'''(c))^2}{24(F''(c))^3}\right]\right\}=\\
\label{eqS5}=&\sqrt{\dfrac{2\pi}{-NF''(c)}}e^{NF_N(c_N)}\left[\left.\left(\dfrac{\partial p^{(0)}(\bar{x})}{\partial h}\right)^2\right|_{\bar{x}=\bar{c}}+\dfrac{A(J,h)}{N}\right].
\end{align}
As $N\rightarrow\infty$, the numerator of the second fraction in $\eqref{eqS3}$ can be approximated as:
\begin{align}
\nonumber &\left(\sqrt{\dfrac{2\pi}{-NF''(c)}}e^{NF_N(c_N)}\left\{\left.\dfrac{\partial p^{(0)}(\bar{x})}{\partial h}\right|_{\bar{x}=\bar{c}}+\dfrac{1}{N}\left[\dfrac{-\dfrac{\text{d}^2}{\text{d} x^2}\left.\dfrac{\partial p^{(0)}(\bar{c})}{\partial h}\right|_{\bar{x}=\bar{c}}}{2F''(c)}+\right.\right.\right.\\
\nonumber&+\left.\left.\left.\dfrac{\left.\dfrac{\partial p^{(0)}(\bar{x})}{\partial h}\right|_{\bar{x}=\bar{c}}F^{(iv)}(c)}{8(F''(c))^2}+\dfrac{\dfrac{\text{d}}{\text{d} x}\left(\left.\dfrac{\partial p^{(0)}(\bar{x})}{\partial h}\right|_{\bar{x}=\bar{c}}\right)^2F'''(c)}{2(F''(c))^2}-\dfrac{5\left.\dfrac{\partial p^{(0)}(\bar{x})}{\partial h}\right|_{\bar{x}=\bar{c}}(F'''(c))^2}{24(F''(c))^3}\right]\right\}\right)^2=\\
\nonumber=&\left(\sqrt{\dfrac{2\pi}{-NF''(c)}}e^{NF_N(c_N)}\left[\left.\dfrac{\partial p^{(0)}(\bar{x})}{\partial h}\right|_{\bar{x}=\bar{c}}+\dfrac{B(J,h)}{N}\right]\right)^2=
\\
\label{eqS6}=&\left(\sqrt{\dfrac{2\pi}{-NF''(c)}}e^{NF_N(c_N)}\right)^2\left[\left.\left(\dfrac{\partial p^{(0)}(\bar{x})}{\partial h}\right)^2\right|_{\bar{x}=\bar{c}}+2\left.\dfrac{\partial p^{(0)}(\bar{x})}{\partial h}\right|_{\bar{x}=\bar{c}}\dfrac{B(J,h)}{N}+\dfrac{B^2(J,h)}{N^2}\right].
\end{align}
As $N\rightarrow\infty$, the integral $\displaystyle\int_{\mathbb{R}}e^{NF_N(x)}dx$ can be approximated as:
\begin{align}
\nonumber&\sqrt{\dfrac{2\pi}{-NF''(c)}}e^{NF_N(c_N)}\left\{1+\dfrac{1}{N}\left[\dfrac{F^{(iv)}(c)}{8(F''(c))^2}-\dfrac{5(F'''(c))^2}{24(F''(c))^3}\right]\right\}=\\
\label{eqS7}=&\sqrt{\dfrac{2\pi}{-NF''(c)}}e^{NF_N(c_N)}\left[1+\dfrac{C(J,h)}{N}\right].
\end{align}
Putting together $\eqref{eqS4}$ and $\eqref{eqS7}$ we obtain:
\begin{equation}\label{eqS8}
\eqref{eqS2}\xrightarrow{N\rightarrow\infty} g'(\bar{x}).\end{equation}
Putting together $\eqref{eqS5}$,$\eqref{eqS6}$ and $\eqref{eqS7}$, we obtain:
\begin{equation}\label{eqS9}
\eqref{eqS3}\xrightarrow{N\rightarrow\infty} \dfrac{-8J^2(g'(\bar{c}))^2}{2(-2J+4J^2g'(\bar{c}))}.\end{equation}
Using \eqref{eqS8} and \eqref{eqS9}, we find that as $N\rightarrow\infty$
\begin{align}
\chi_N(J,h)\xrightarrow{N\rightarrow\infty}g'+\dfrac{8J^2(g')^2}{2(-2J+4J^2g')}=\dfrac{4Jg'(1-2Jg')+8J^2(g')^2}{4J(1-2Jg')}=\dfrac{(g')^2}{1-2Jg'}.
\end{align}
At the thermodynamic limit, the susceptibility is the partial derivative of the solution $m(J,h)$ of the consistency equation with respect to the parameter $h$, so that: 
\begin{align*}
\chi&=\dfrac{\partial m(J,h)}{\partial h}=\dfrac{\text{d}}{\text{d} h}g((2m-1)J+h)\left(1+2\dfrac{\partial m(J,h)}{\partial h}J\right)=\\
&=g'((2m-1)J+h)(1+2\chi J)\\
\Rightarrow \chi&=\dfrac{(g')^2}{1-2Jg'}.
\end{align*}
Hence, \eqref{eqS1} is proved.
\endproof
\end{thm}

\section{Extended Laplace's method. Control at the second order.}
The usual Laplace method works with integrals of the form
$$\int_{\mathbb{R}}(\psi(x))^n u(x) dx$$
as $n\rightarrow\infty$. In this appendix we prove an extension at the second order of the previous method when the functions $\psi$ and $u$ may depend on $n$ (see \cite{ACFM14} for the control at first order). We have used that in Appendix A.
\begin{thm}
For all $n\in\mathbb{N},$ let $\psi_n:\mathbb{R}\rightarrow\overline{\mathbb{R}}$ and $u_n:\mathbb{R}\rightarrow\overline{\mathbb{R}}$.
Suppose that there exists a compact interval $K\subset\mathbb{R}$ such that $\psi_n,u_n>0$ on $K$, so that in particular
$$\psi_n(x)=e^{f_n(x)}\quad \forall x\in K.$$
Suppose that $f_n\in C^4(K)$ and that $u_n\in C^2(K)$. \\
Moreover suppose that
\begin{enumerate}
\item $f_n\xrightarrow{n\rightarrow\infty}f$ uniformly on $K$ with its derivatives;
\item $u_n\xrightarrow{n\rightarrow\infty}u$ uniformly on $K$ with its derivatives;
\item there exixts a positive constant $c_1<\infty$ such that $|u_n|<c_1$;
\item $\max\limits_K f_n$ is attained in a unique point $c_n\in \text{int}(K)$;
\item $\max\limits_{K}f$ is attained in a unique point $c\in \text{int}(K)$;
\item $\limsup\limits_{n\rightarrow\infty}\left(\sup\limits_{\mathbb{R}\setminus K}\log|\psi_n|-\max\limits_{K}f_n\right)<0$;
\item $f''(c)<0$;
\item $\limsup\limits_{n\rightarrow\infty}\int_{\mathbb{R}}|\psi_n(x)|dx<\infty$.
\end{enumerate} 
Then, as $n\rightarrow\infty$,
\begin{align}\label{eqL1}
\int_{\mathbb{R}}(\psi_n(x))^n u_n(x) dx = \sqrt{\dfrac{2\pi}{-nf''(c)}}e^{nf_n(c_n)}\left\{u(c)+\dfrac{\Lambda}{n}+o\left(\dfrac{1}{n}\right)\right\},
\end{align}
where $$\Lambda=-\dfrac{u''(c)}{2f''(c)}+\dfrac{u(c)f^{(iv)}(c)}{8(f''(c))^2}+\dfrac{u'(c)f'''(c)}{2(f''(c))^2}-\dfrac{5u(c)(f'''(c))^2}{24(f''(c))^3}.$$
\end{thm}
In the proof we use the following elementary fact:
\begin{lem}\label{lemL1}
Let $(f_n)_{n\in\mathbb{N}}$ be a sequence of continuous functions uniformly convergent to $f$ on a compact set $K$. Let $(I_n)_{n\in\mathbb{N}}$ and $I$ be subsets of $K$ such that $$\max\limits_{x\in I_n,y\in I}\text{dist}(x,y)\rightarrow 0,\quad\text{as}\quad n\rightarrow\infty.$$
Then
\begin{itemize}
\item[a)]$\max\limits_{I_n}f_n\xrightarrow{n\rightarrow\infty}\max\limits_{I}f$
\item[b)]$\text{arg}\max\limits_{I_n}f_n\xrightarrow{n\rightarrow\infty}\text{arg}\max\limits_{I}f$, provided that $f$ has a unique global maximum point on $I$.
\end{itemize}
\end{lem}
We proceed with the proof of the theorem.
\proof
Since $c_n$ is an internal point of maximum of $f_n$ (\emph{hypothesis 4}), $f_n'(c_n)=0$. Moreover $\forall x \in K$
\begin{align}\label{eqL2}
f_n(x)=f_n(c_n)+\dfrac{1}{2}f_n''(c_n)(x-c_n)^2+\dfrac{1}{6}f_n'''(c_n)(x-c_n)^3+\dfrac{1}{24}f_n^{(iv)}(\xi'_{x,n})(x-c_n)^4,
\end{align}
with $\xi'_{x,n}\in(c_n,x)\subset K$,
and
\begin{align}\label{eqL6bis}
u_n(x)=u_n(c_n)+u_n'(c_n)(x-c_n)+\dfrac{1}{2}u_n''(\xi''_{x,n})(x-c_n)^2,
\end{align}
with $\xi''_{x,n}\in(c_n,x)\subset K.$\\
Fix $\dfrac{\epsilon}{2}=\overline{\epsilon}$ and $N_{\epsilon}$ such that $|f_n^{(i)}(\xi)-f^{(i)}(\xi)|<\overline{\epsilon},i=1,2,3,4$ and $|u_n^{(j)}(\xi)-u^{(j)}(\xi)|<\overline{\epsilon},j=1,2$ $\forall\xi\in K$ and $\forall n>N_{\epsilon}$.
Since $f$ and $u$ and their respective derivatives are continuous in $c$, there exists $\overline{\delta}_{\epsilon}>0$ such that $B(c,\overline{\delta}_{\epsilon})\subset K$ and $\forall\xi:|\xi-c|<\overline{\delta}_{\epsilon}$
\begin{align*}
|f^{(i)}(\xi)-f^{(i)}(c)|&<\dfrac{\epsilon}{2},i=1,2,3,4,
\end{align*}
and 
\begin{align*}
|u^{(j)}(\xi)-u^{(j)}(c)|&<\dfrac{\epsilon}{2},j=1,2.
\end{align*}
By lemma \ref{lemL1}, $c_n\xrightarrow{n\rightarrow\infty}c$ because $c$ is the unique maximum point of $f$ on $K$(\emph{hypothesis 5}). Thus there exists $\overline{N}_{\delta_{\epsilon}}$ such that
\begin{equation}
|c_n-c|<\delta_{\epsilon}=\dfrac{\overline{\delta}_{\epsilon}}{3}\quad \forall n>\overline{N}_{\delta_{\epsilon}}.
\end{equation}
Observe that by \emph{hypothesis 7} and for $n>N_{\epsilon}\vee \overline{N}_{\delta_{\epsilon}}$, $f_n''(c_n)<0.$\\
Moreover, for $ n>N_{\epsilon}\vee \overline{N}_{\delta_{\epsilon}}$, $\forall x\in B(c,\delta_{\epsilon})\subset B(c,\overline{\delta}_{\epsilon})$ and $\forall\xi_{x,n}\in(c_n,x)\subset K$, it holds:
\begin{align}
\nonumber &|\xi_{x,n}-c|\leq|\xi_{x,n}-x|+|x-c|\leq|c_n-x|+|x-c|\leq|c_n-c|+|c-x|+|x-c|<3\delta_{\epsilon}=\overline{\delta}_{\epsilon}\Rightarrow
\end{align}
\begin{align}\label{eqL2ter}
\begin{cases}
&|f_n^{(i)}(\xi'_{x,n})-f^{(i)}(c)|\leq|f_n^{(i)}(\xi'_{x,n})-f^{(i)}(\xi'_{x,n})|+|f^{(i)}(\xi'_{x,n})-f^{(i)}(c)|<\dfrac{\epsilon}{2}+\dfrac{\epsilon}{2}=\epsilon\\
&|u_n^{(j)}(\xi''_{x,n})-u^{(j)}(c)|\leq|u_n^{(j)}(\xi''_{x,n})-u^{(j)}(\xi''_{x,n})|+|u^{(i)}(\xi''_{x,n})-u^{(i)}(c)|<\dfrac{\epsilon}{2}+\dfrac{\epsilon}{2}=\epsilon
\end{cases}
\end{align}
By substituing \eqref{eqL2ter} in \eqref{eqL2} and in \eqref{eqL6bis}, we obtain that for $n>N_{\epsilon}\vee \overline{N}_{\delta_{\epsilon}}$ and $x\in B(c,\delta_{\epsilon})$
\begin{align}\label{eqL2bis1}
f_n(x)
\begin{cases}
\leq &f_n(c_n)+\dfrac{1}{2}f_n''(c_n)(x-c_n)^2+\dfrac{1}{6}f_n'''(c_n)(x-c_n)^3+\\
&+\dfrac{1}{24}(f^{(iv)}(c)+\epsilon)(x-c_n)^4\\
\geq &f_n(c_n)+\dfrac{1}{2}f_n''(c_n)(x-c_n)^2+\dfrac{1}{6}f_n'''(c_n)(x-c_n)^3+\\
&+\dfrac{1}{24}(f^{(iv)}(c)-\epsilon)(x-c_n)^4
\end{cases}
\end{align}
and
\begin{align}\label{eqL2bis2}
u_n(x)
\begin{cases}
\leq &u_n(c_n)+u_n'(c_n)(x-c_n)+\dfrac{1}{2}(u''(c)+\epsilon)(x-c_n)^2\\
\geq &u_n(c_n)+u_n'(c_n)(x-c_n)+\dfrac{1}{2}(u''(c)-\epsilon)(x-c_n)^2
\end{cases}
.
\end{align}
Now split the integral into two parts:
\begin{equation}\label{eqL3}
\int_{\mathbb{R}}e^{nf_n(x)}u_n(x) dx=\int_{\mathbb{R}\setminus B(c_n,\delta_{\epsilon})}(\psi_n(x))^nu_n(x) dx+\int_{B(c_n,\delta_{\epsilon})}e^{nf_n(x)}u_n(x) dx.
\end{equation}
To control the first integral on the r.h.s. of \eqref{eqL3} we claim that there exists $\eta_{\delta_{\epsilon}}>0$ and $\hat{N}_{\delta_{\epsilon}}$ such that
\begin{equation}\label{eqL4}
\log|\psi_n(x)|<f_n(c_n)-\eta_{\delta_{\epsilon}} \quad \forall x\in\mathbb{R}\setminus B(c_n,\delta_{\epsilon})\quad\forall n>\hat{N}_{\delta_{\epsilon}};
\end{equation}
this implies that
$$\limsup\limits_{n\rightarrow\infty}\sup\limits_{x\in\mathbb{R}\setminus B(c_n,\delta_{\epsilon})}\left(\log|\psi_n(x)|-f_n(c_n)\right)<0.$$
Indeed, using lemma \ref{lemL1}:
\begin{align*}
&\limsup\limits_{n\rightarrow\infty}\sup\limits_{\mathbb{R}\setminus B(c_n,\delta_{\epsilon})}\left(\log|\psi_n(x)|-f_n(c_n)\right)=\\
&\left(\limsup\limits_{n\rightarrow\infty}\sup\limits_{x\in K\setminus B(c_n,\delta_{\epsilon})}(f_n(x)-f_n(c_n))\right)\vee\left(\limsup\limits_{n\rightarrow\infty}\sup\limits_{x\in\mathbb{R}\setminus K}(\log|\psi_n(x)|-f_n(c_n)\right)=\\
&\left(\sup\limits_{x\in K\setminus B(c_n,\delta_{\epsilon})}(f(x)-f(c))\right)\vee\left(\limsup\limits_{n\rightarrow\infty}\sup\limits_{x\in\mathbb{R}\setminus K}(\log|\psi_n(x)|-f_n(c_n))\right).
\end{align*}
Moreover, since $c$ is the unique maximum point of the continuous function $f$ on the compact set $K$, $$\sup\limits_{x\in K\setminus B(c_n,\delta_{\epsilon})}(f(x)-f(c))<0$$and this proves the claim.\\
Now using \eqref{eqL4} and \emph{hypothesis 8} we can say that there exist $C_1$ and $N$ such that for all $n>N\vee \hat{N}_{\delta_{\epsilon}}$ 
\begin{align}\label{eqL4bis}
\nonumber\int_{\mathbb{R}\setminus B(c_n,\delta_{\epsilon})}e^{nf_n(x)}u_n(x) dx &\leq e^{(n-1)(f_n(c_n)-\eta\delta_{\epsilon})}\int_{\mathbb{R}}|\psi_n(x)||u_n(x)| dx\leq\\
\nonumber&\leq e^{(n-1)(f_n(c_n)-\eta\delta_{\epsilon})}\int_{\mathbb{R}}|\psi_n(x)|c_1 dx\leq\\
&\leq C_1e^{n(f_n(c_n)-\eta\delta_{\epsilon})}.
\end{align}
In order to find an upper bound for the second integral of the r.h.s. of \eqref{eqL3}, we proceed as follows:
\begin{align}\label{eqL5}
\nonumber\int_{B(c_n,\delta_{\epsilon})}&e^{nf_n(x)}u_n(x) dx\leq\\
\nonumber\leq \int_{B(c_n,\delta_{\epsilon})}&e^{n\left(f_n(c_n)+\frac{1}{2}f_n''(c_n)(x-c_n)^2+\frac{1}{6}f_n'''(c_n)(x-c_n)^3+\frac{1}{24}(f^{(iv)}(c)+\epsilon)(x-c_n)^4\right)}u_n(x)dx=\\
=\int_{B(c_n,\delta_{\epsilon})}&e^{nf_n(c_n)+\frac{n}{2}f_n''(c_n)(x-c_n)^2}e^{n\left(\frac{1}{6}f_n'''(c_n)(x-c_n)^3+\frac{1}{24}(f^{(iv)}(c)+\epsilon)ì(x-c_n)^4\right)}u_n(x)dx.
\end{align}
Since $\delta_{\epsilon}$ may be chosen small, the second exponential term can be expanded as
\begin{align}\label{eqL6}
\nonumber&\exp\left[n\left(\frac{1}{6}f_n'''(c_n)(x-c_n)^3+\frac{1}{24}(f^{(iv)}(c)+\epsilon)(x-c_n)^4\right)\right]\leq\\
\nonumber\leq&1+n\left[\frac{1}{6}f_n'''(c_n)(x-c_n)^3+\frac{1}{24}(f^{(iv)}(c)+\epsilon)(x-c_n)^4\right]+\\
&+\dfrac{n^2}{72}(f_n'''(c_n))^2(x-c_n)^6+n^2C_2|x-c_n|^7,
\end{align}
where $C_2$ is a positive real constant. 
Substitute \eqref{eqL6} and \eqref{eqL2bis2} in \eqref{eqL5}. Collecting powers of $(x-c_n)$ and observing that odd powers don't contribute to the integral, we claim that:
\begin{align*}
\eqref{eqL5}\leq&e^{nf_n(c_n)}\int_{B(c_n,\delta_{\epsilon})}e^{\frac{n}{2}f_n''(c_n)(x-c_n)^2}\left[u_n(c_n)+
(x-c_n)^2\dfrac{u''(c)+\epsilon}{2}+\right.\\
&+(x-c_n)^4\left(n\dfrac{u_n'(c_n)f_n'''(c_n)}{6}+n\dfrac{u_n(c_n)(f^{(iv)}(c)+\epsilon))}{24}\right)+\\
&+(x-c_n)^6\left(n^2\dfrac{u_n(c_n)(f_n'''(c_n))^2}{72}+n\dfrac{(u''(c)+\epsilon)(f^{(iv)}(c)+\epsilon)}{48}\right)+\\
&\left.+(x-c_n)^8n^2\left(\dfrac{(u''(c)+\epsilon)(f_n'''(c_n))^2}{144}+u_n'(c_n)C_2\right)\right]dx+\\
+&e^{nf_n(c_n)}\int_{B(c_n,\delta_{\epsilon})}e^{\frac{n}{2}f_n''(c_n)(x-c_n)^2}n^2C_2|x-c_n|^7dx.
\end{align*}
Making the change of variable $$t=\sqrt{-nf_n''(c_n)}(x-c_n),$$ we obtain: 
\begin{align}\label{eqL7}
\nonumber\eqref{eqL5}\leq&\dfrac{e^{nf_n(c_n)}}{\sqrt{-nf_n''(c_n)}}\int_{B(0,\sqrt{-nf_n''(c_n)}\delta_{\epsilon})}e^{-\frac{t^2}{2}}\left\{u_n(c_n)+\dfrac{1}{n}\left[-t^2\dfrac{u''(c)+\epsilon}{2f_n''(c_n)}+\right.\right.\\
\nonumber&\left.+t^4\left(\frac{u_n(c_n)(f^{(iv)}(c)+\epsilon)}{24(f_n''(c_n))^2}+\frac{u_n'(c_n)f_n'''(c_n)}{6(f_n''(c_n))^2}\right)-t^6\frac{u_n(c_n)(f_n'''(c_n))^2}{72(f_n''(c_n))^3}\right]+\\
\nonumber&\left.+\dfrac{1}{n^2}\left[-t^6\dfrac{(u''(c)+\epsilon)(f^{(iv)}(c)+\epsilon)}{48(f_n''(c_n))^3}+t^8\left(\frac{(u''(c)+\epsilon)(f_n'''(c_n))^2}{144(f_n''(c_n))^4}+\frac{u_n'(c_n)C_2}{(f_n''(c_n))^4}+\right)\right]\right\}dt\\
\nonumber&+2\dfrac{e^{nf_n(c_n)}}{\sqrt{-nf_n''(c_n)}}\int_{t\in B(0,\sqrt{-nf_n''(c_n)}\delta_{\epsilon}):t\geq0}e^{-\frac{t^2}{2}}t^7\dfrac{C_2}{n^{3/2}(f_n''(c_n))^{7/2}}dt=\\
\nonumber=&\dfrac{e^{nf_n(c_n)}}{\sqrt{-n(f_n''(c_n))}}\int_{B(0,\sqrt{-nf_n''(c_n)}\delta_{\epsilon})}e^{-\frac{t^2}{2}}\left\{u_n(c_n)+\dfrac{a^{(1)}_{n,\epsilon}(t)}{n}+\dfrac{b^{(1)}_{n,\epsilon}(t)}{n^2}\right\}dt+\\
&+2\dfrac{C_2e^{nf_n(c_n)}}{n^2(f_n''(c_n))^4}\int_{t\in B(0,\sqrt{-n(f_n''(c_n))}\delta_{\epsilon}):t\geq0}e^{-\frac{t^2}{2}}t^7dt,
\end{align}
where $a^{(1)}_{n,\epsilon}(t)$ and $b^{(1)}_{n,\epsilon}(t)$ are the arguments inside square brackets which are respectively multiplied by $\frac{1}{n}$ and by $\frac{1}{n^2}$.\\\\
In order to find a lower bound for the second integral of the r.h.s. of \eqref{eqL3}, we proceed as follows:
\begin{align}\label{eqL8}
\nonumber&\int_{\mathbb{R}}(\psi_n(x))^n u_n(x) dx \geq\\
\geq&\int_{B(c_n,\delta_{\epsilon})}e^{nf_n(c_n)+\frac{n}{2}f_n''(c_n)(x-c_n)^2}e^{n\left(\frac{1}{6}f_n'''(c_n)(x-c_n)^3+\frac{1}{24}(f^{(iv)}(c)-\epsilon)ì(x-c_n)^4\right)}u_n(x)dx.
\end{align}
Since $\delta_{\epsilon}$ may be chosen small, the second exponential term satisfies
\begin{align}\label{eqL9}
\nonumber&\exp\left[n\left(\frac{1}{6}f_n'''(c_n)(x-c_n)^3+\frac{1}{24}(f^{(iv)}(c)-\epsilon)(x-c_n)^4\right)\right]\geq\\
\nonumber\geq&1+n\left[\frac{1}{6}f_n'''(c_n)(x-c_n)^3+\frac{1}{24}(f^{(iv)}(c)-\epsilon)(x-c_n)^4\right]+\\
&+\dfrac{n^2}{72}(f_n'''(c_n))^2(x-c_n)^6-n^2C_3|x-c_n|^7,
\end{align}
where $C_3$ is a positive real constant.\\
Analogously as above, expand the second exponential term of \eqref{eqL8} as in \eqref{eqL6} and the function $u_n(x)$ as in \eqref{eqL6bis}.
Collecting powers of $(x-c_n)$ and making the change of variable $$t=\sqrt{-nf_n''(c_n)}(x-c_n),$$ we obtain:
\begin{align}\label{eqL10}
\nonumber\eqref{eqL8}\geq&\dfrac{e^{nf_n(c_n)}}{\sqrt{-nf_n''(c_n)}}\int_{B(0,\sqrt{-nf_n''(c_n)}\delta_{\epsilon})}e^{-\frac{t^2}{2}}\left\{u_n(c_n)+\dfrac{1}{n}\left[-t^2\dfrac{u''(c)-\epsilon}{2f_n''(c_n)}+\right.\right.\\
\nonumber&\left.+t^4\left(\frac{u_n(c_n)(f^{(iv)}(c)-\epsilon)}{24(f_n''(c_n))^2}+\frac{u_n'(c_n)f_n'''(c_n)}{6(f_n''(c_n))^2}\right)-t^6\frac{u_n(c_n)(f_n'''(c_n))^2}{72(f_n''(c_n))^3}\right]+\\
\nonumber&+\left.\dfrac{1}{n^2}\left[-t^6\dfrac{(u''(c)-\epsilon)(f^{(iv)}(c)-\epsilon)}{48(f_n''(c_n))^3}+t^8\left(\frac{(u''(c)-\epsilon)(f_n'''(c_n))^2)}{144(f_n''(c_n))^4}-\frac{u_n'(c_n)C_3}{(f_n''(c_n))^4}\right)\right]\right\}dt+\\
\nonumber&-2\dfrac{e^{nf_n(c_n)}}{\sqrt{-nf_n''(c_n)}}\int_{t\in B(0,\sqrt{-nf_n''(c_n)}\delta_{\epsilon}):t\geq0}e^{-\frac{t^2}{2}}t^7\dfrac{C_3}{n^{3/2}(f_n''(c_n))^{7/2}}dt=\\
\nonumber=&\dfrac{e^{nf_n(c_n)}}{\sqrt{-nf_n''(c_n)}}\int_{B(0,\sqrt{-nf_n''(c_n)}\delta_{\epsilon})}e^{-\frac{t^2}{2}}\left\{u_n(c_n)+\dfrac{a^{(2)}_{n,\epsilon}(t)}{n}+\dfrac{b^{(2)}_{n,\epsilon}(t)}{n^2}\right\}dt+\\
&-2\dfrac{C_3e^{nf_n(c_n)}}{n^2(f_n''(c_n))^4}\int_{t\in B(0,\sqrt{-n(f_n''(c_n))}\delta_{\epsilon}):t\geq0}e^{-\frac{t^2}{2}}t^7dt,
\end{align}
where $a^{(2)}_{n,\epsilon}(t)$ and $b^{(2)}_{n,\epsilon}(t)$ are the arguments inside square brackets which are respectively multiplied by $\frac{1}{n}$ and by $\frac{1}{n^2}$.\\\\
%
It is easy to verify that:
\begin{align}\label{eqL11}
\nonumber&\int_{B(0,\sqrt{n}\delta_{\epsilon})}e^{-\frac{t^2}{2}}u_n(c_n)dt\xrightarrow{n\rightarrow{\infty}}\sqrt{2\pi}u(c),\\
\nonumber&\int_{B(0,\sqrt{n}\delta_{\epsilon})}e^{-\frac{t^2}{2}}t^{2k}dt\xrightarrow{n\rightarrow{\infty}}\int_{\mathbb{R}}e^{-\frac{t^2}{2}}t^{2k}dt=\sqrt{2\pi}(2k-1)(2k-3)\ldots(3)(1),\quad\forall k\in\mathbb{N},\\
&\int_{t\in B(0,\sqrt{n}\delta_{\epsilon}):t\geq0}e^{-\frac{t^2}{2}}t^7dt\xrightarrow{n\rightarrow{\infty}}\int_0^{+\infty}e^{-\frac{t^2}{2}}t^7dt=48.
\end{align}\\
In conclusion, using \eqref{eqL3},\eqref{eqL4bis},\eqref{eqL7}, \eqref{eqL10} and \eqref{eqL11}, we obtain that for $\epsilon\in(0,\epsilon_0]$ and $n>N\vee N_{\epsilon}\vee\overline{N}_{\delta_{\epsilon}}\vee\hat{N}_{\delta_{\epsilon}}$
\begin{align}\label{eqL12}
\nonumber&\dfrac{\displaystyle\int_{\mathbb{R}}(\psi_n(x))^n u_n(x) dx-\sqrt{\dfrac{2\pi}{-nf''(c)}}e^{nf_n(c_n)}u(c)}{\sqrt{\dfrac{2\pi}{-nf''(c)}}e^{nf_n(c_n)}\dfrac{\Lambda}{n}}\leq\\
\nonumber\leq&\dfrac{\dfrac{\displaystyle\int_{B(0,\sqrt{-nf_n''(c_n)}\delta_{\epsilon})}e^{-\frac{t^2}{2}}\left\{u_n(c_n)+\dfrac{a^{(1)}_{n,\epsilon}(t)}{n}+\dfrac{b^{(1)}_{n,\epsilon}(t)}{n^2}\right\}dt}{\sqrt{-f_n''(c_n)}}-\sqrt{\dfrac{2\pi}{-nf''(c)}}u(c)}{\sqrt{\dfrac{2\pi}{-f''(c)}}\dfrac{\Lambda}{n}}+\\
\nonumber&+\dfrac{2\dfrac{C_2}{n^2(f_n''(c_n))^5}\displaystyle\int_{t\in B(0,\sqrt{-nf_n''(c_n)}\delta_{\epsilon}):t\geq0}e^{-\frac{t^2}{2}}t^7dt+\dfrac{C_1}{e^{n\eta\delta_{\epsilon}}}}{\sqrt{\dfrac{2\pi}{-f''(c)}}\dfrac{\Lambda}{n}}\xrightarrow{n\rightarrow\infty}\\
&\xrightarrow{n\rightarrow\infty}\dfrac{\sqrt{\dfrac{2\pi}{-f_n''(c_n)}}}{\sqrt{\dfrac{2\pi}{-f''(c)}}}\xrightarrow{\epsilon\rightarrow0}1
\end{align}
and
\begin{align}\label{eqL13}
\nonumber&\dfrac{\displaystyle\int_{\mathbb{R}}(\psi_n(x))^n u_n(x) dx-\sqrt{\dfrac{2\pi}{-nf''(c)}}e^{nf_n(c_n)}u(c)}{\sqrt{\dfrac{2\pi}{-nf''(c)}}e^{nf_n(c_n)}\dfrac{\Lambda}{n}}\geq\\
\nonumber\geq&\dfrac{\dfrac{\displaystyle\int_{B(0,\sqrt{nf_n''(c_n)}\delta_{\epsilon})}e^{-\frac{t^2}{2}}\left\{u_n(c_n)+\dfrac{a^{(2)}_{n,\epsilon}(t)}{n}+\dfrac{b^{(2)}_{n,\epsilon}(t)}{n^2}\right\}dt}{\sqrt{-f_n''(c_n)}}-\sqrt{\dfrac{2\pi}{-nf''(c)}}u(c)}{\sqrt{\dfrac{2\pi}{-f''(c)}}\dfrac{\Lambda}{n}}+\\
\nonumber&-\dfrac{2\dfrac{C_3}{n^2(f_n''(c_n))^5}\displaystyle\int_{t\in B(0,\sqrt{-nf_n''(c_n)}\delta_{\epsilon}):t\geq0}e^{-\frac{t^2}{2}}t^7dt+\dfrac{C_1}{e^{n\eta\delta_{\epsilon}}}}{\sqrt{\dfrac{2\pi}{-f''(c)}}\dfrac{\Lambda}{n}}\xrightarrow{n\rightarrow\infty}\\
&\xrightarrow{n\rightarrow\infty}\dfrac{\sqrt{\dfrac{2\pi}{-f_n''(c_n)}}}{\sqrt{\dfrac{2\pi}{-f''(c)}}}\xrightarrow{\epsilon\rightarrow0}1.
\end{align}
Hence, by \eqref{eqL12} and \eqref{eqL13}, \eqref{eqL1} is proved.
\endproof

{\bf Acknowledgements.} The authors wish to thank Diego Alberici, Claudio Giberti and Emanuele Mingione for interesting discussions.
This work was partially supported by PRIN Grant N. 2010HXAW77-010: {\it Statistical Mechanics of disordered and complex systems},
and FIRB Grant N. RBFR10N90W: {\it Stochastic Processes in Interacting Particle Systems: Duality, Metastability and their Applications.}


\end{document}